\documentclass[aps,reprint,pra,superscriptaddress,floats,floatfix,nobibnotes]{revtex4-2}
\usepackage{amsmath}
\usepackage{amsfonts}
\usepackage{amssymb}
\usepackage{graphicx}
\usepackage{color}
\usepackage[colorlinks=true,citecolor=blue,linkcolor=blue,urlcolor=blue]{hyperref}
\usepackage{times}
\usepackage{amstext}
\usepackage{latexsym}
\usepackage{dcolumn}
\usepackage{bm}
\usepackage{mathtools}

\providecommand{\moy}[1]{\langle #1 \rangle}
\DeclarePairedDelimiter\bra{\langle}{\rvert}
\DeclarePairedDelimiter\ket{\lvert}{\rangle}
\DeclarePairedDelimiterX\braket[2]{\langle}{\rangle}{#1 \delimsize\vert #2}

\begin{document}

\title{Universal quantum computation using atoms in cross-cavity systems}

\author{Luiz O. R. Solak}
\email{solakluiz@estudante.ufscar.br}
\affiliation{Departamento de Física, Universidade Federal de São Carlos, 13565-905 São Carlos, São Paulo, Brazil}
\author{Daniel Z. Rossatto}
\email{dz.rossatto@unesp.br}
\affiliation{Universidade Estadual Paulista (UNESP), Instituto de Ciências e Engenharia, 18409-010 Itapeva, São Paulo, Brazil}
\author{Celso J. Villas-Boas}
\email{celsovb@df.ufscar.br}
\affiliation{Departamento de Física, Universidade Federal de São Carlos, 13565-905 São Carlos, São Paulo, Brazil}

\begin{abstract}
Quantum gates are the building blocks of quantum circuits, which in turn are the cornerstones of quantum information processing. In this work, we theoretically investigate a single-step implementation of both a universal two- (CNOT) and three-qubit (quantum Fredkin) gates in a cross-cavity setup coupled to a $\Lambda$-type three-level atom. Within a high-cooperativity regime, the system exhibits an atomic-state-dependent $\pi$-phase gate involving the two-mode single-photon bright and dark states of the input light pulses. This allows for the controlled manipulation of light states by the atom and vice versa. Our results indicate these quantum gates can be implemented with high probability of success using the state-of-the-art parameters, either for the weak- or strong-coupling regime, where the quantum interference is due to an electromagnetically-induced-transparency-like phenomenon and the Autler-Townes splitting, respectively. This work not only paves the way for implementing quantum gates in a single step using simple resources, thus avoiding the need to chain basic gates together in a circuit, but it also endorses the potential of cross-cavity systems for realizing universal quantum computation.
\end{abstract}

\maketitle

\section{Introduction}
Efficient manipulation of quantum information is crucial for the realization of quantum computers~\cite{Ladd2010}, which can improve the performance of computing tasks~\cite{Monz2016} or even solve problems that are intractable for classical computers~\cite{Zhong2020}. A quantum computation is carried out through a quantum circuit composed of a sequence of quantum operations, such as quantum gates, measurements, and initialization of qubits~\cite{chuang}. However, the implementation of a multi-qubit gate in the standard circuit model remains challenging because it needs to be decomposed into a chain of universal quantum gates, comprising single- and two-qubit basic gates~\cite{chuang}. For example, the three-qubit Fredkin gate~\cite{PhysRevA.53.2855} is fundamental for universal reversible computation as it plays important roles in quantum computing~\cite{Vandersypen2001,MartnLpez2012,PhysRevLett.99.250505}, quantum cryptography~\cite{PhysRevLett.87.167902,PhysRevLett.95.150502}, quantum error correction~\cite{PhysRevLett.76.4281,Barenco1997} and measurement~\cite{PhysRevLett.88.217901,PhysRevLett.89.190401}. However, its implementation requires at least five two-qubit gates~\cite{PhysRevA.53.2855}, then it is of interest to implement this gate more directly~\cite{Patel2016,Li2022}, without decomposition, in order to mitigate the resource overhead, thus enhancing the overall probability of success to implement it.
Various physical systems are currently being explored as potential candidates for the implementation of large-scale universal quantum computing~\cite{Vandersypen2001,Gong2021, Carolan2015, Ebadi2021, rempe2015}. In particular, we are interested in cavity-based quantum networks with single atoms in optical cavities~\cite{rempe2015,zoller95,cirac97,duan04,kimbleqi,monroerev,Ritter_2012,brekenfeld}, in which a variety of applications have been proposed~\cite{scarani09,qrepeater,kimble08qm,Langenfeld2020,PhysRevLett.126.130502,tiarks2014qt,turchette95,duan05,zhou2011,cvb16,cvbpra18,Hacker2016,Daiss2021,PhysRevX.12.021035,Holger2003}. 

\begin{figure*}[t]
\includegraphics[width =.7\textwidth]{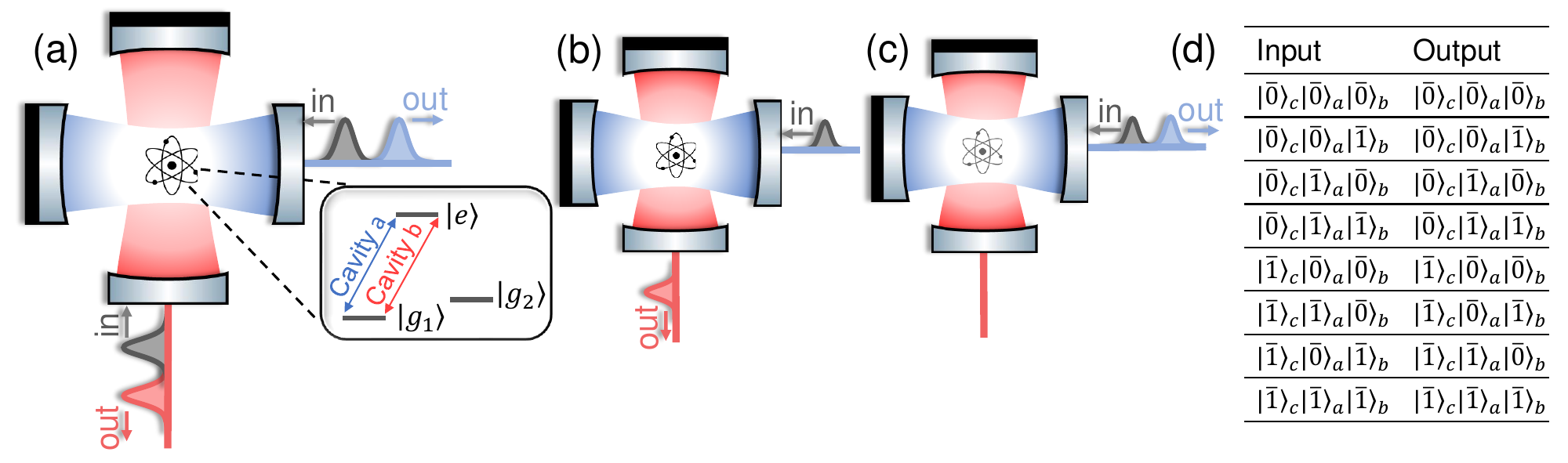}
\includegraphics[trim = 0.1mm 5mm 0mm 5mm, width = 0.25\textwidth, clip]{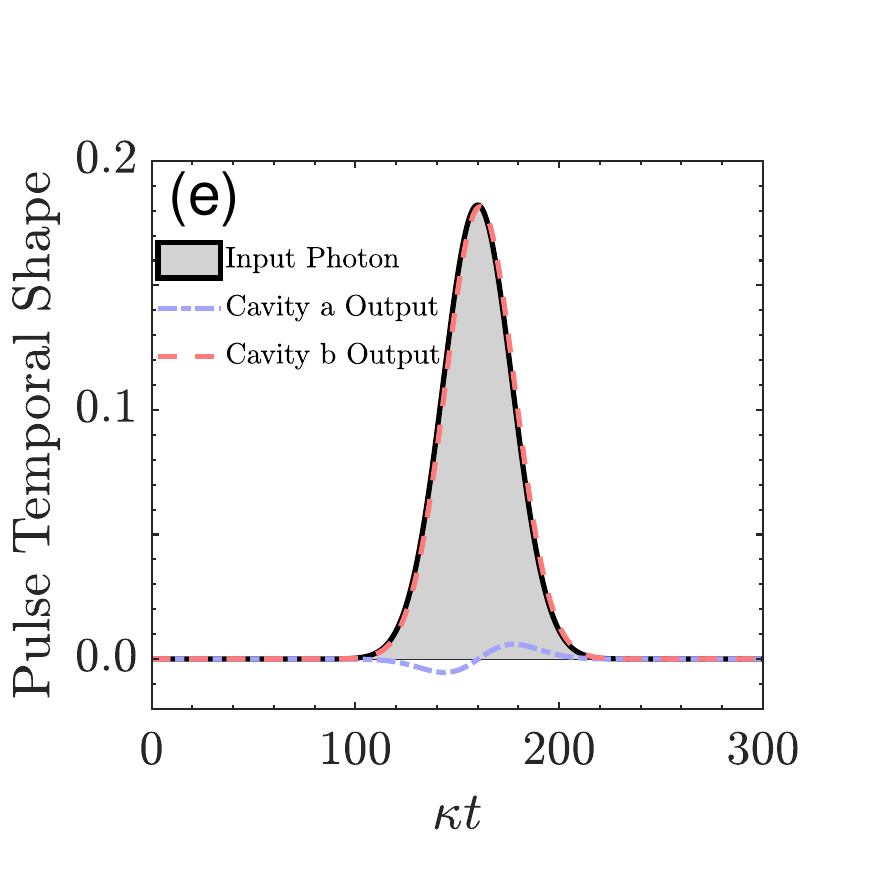}
\caption{\label{scheme1}(a) A $\Lambda$-type three-level atom is resonantly coupled to a cross-cavity system, with single-sided cavities. (b) Impinging a single-photon pulse upon a cavity when the atom is in $\ket{g_1}$, a quantum interference exchanges the cavity light states, resulting in a single-photon pulse transmitted by the opposite cavity. (c) If the atom is is $\ket{g_2}$, it becomes transparent to the cavities, and thus the output single-photon pulse exits from the same cavity through which it entered. (d) Truth table of the Fredkin gate implemented by our system, where the control qubit is given by the atomic ground states ($\ket{g_1}\to\ket{\bar{1}}_{c}$ and $\ket{g_2}\rightarrow\ket{\bar{0}}_{c}$), while the target qubits are given by the vacuum and single-photon states of the reservoirs of cavity $a$ and b: $\ket{\bar{0}}_{a}$, $\ket{\bar{1}}_{a}$ and $\ket{\bar{0}}_{b}$, $\ket{\bar{1}}_{b}$. (e) Dynamics when a single-photon pulse of temporal shape $\alpha_\text{in}(t)$ impinges upon the cavity $a$, but leaves from the cavity $b$ [$\beta_\text{out}(t)$] since the atom is in $\ket{g_1}$. The blue line [$\alpha_\text{out}(t)$] indicates the possibility of the input photon emerging from the wrong output port of the cavity system.}
\end{figure*}

Here, we show a single-step implementation of universal two- (CNOT) and three-qubit (quantum Fredkin) gates in a cross-cavity setup~\cite{brekenfeld}, with single-sided cavities coupled to a three-level atom [Fig.~\ref{scheme1}(a)]. These gates can be implemented with a success probability greater than 95\% (CNOT) and 90\% (Fredkin) considering the state-of-the-art parameters. The principle behind the implementation of these gates is an extension of the Duan-Kimble protocol~\cite{duan04}, where a $\pi$-phase shift acquired by the two-mode input pulse is conditioned not only on the atomic state but also on whether the input field is in a collective dark or bright state. For the CNOT gate, compared to the Duan-Kimble scheme, the probability of failure is around 50\% lower in our case due to the use of collective effects of light.

We consider a $\Lambda$-type three-level atom (two metastable ground states, $\ket{g_1}$ and $\ket{g_2}$, and an excited one $\ket{e}$) trapped inside the cavity setup. The atomic transition $\ket{g_{1}} \leftrightarrow \ket{e}$ is resonantly coupled to both intracavity modes, while the atomic ground state $\ket{g_{2}}$ remains uncoupled. When the atom is in $\ket{g_2}$ or the input field is in a dark state, the atom becomes transparent to the cavity modes, and thus the pulse enters and leaves the cavities without changing its phase and shape (as long as the pulse duration is much longer than the inverse of the cavities linewidth). On the other hand, in a high-cooperativity regime, when the atom is in $\ket{g_1}$ and the input field is in a bright state, the pulse acquires a $\pi$ phase per excitation since it gets reflected by the cavities due to the Autler-Townes splitting (strong atom-cavity coupling) or an electromagnetically-induced-transparency-like phenomenon (weak coupling). Specifically, if the atom is in $\ket{g_1}$ and a single-photon pulse impinges upon the cavity $a$ (or $b$), the light states between the cavities are exchanged due to the $\pi$-phase shift acquired by its bright component, therefore resulting in an output single-photon pulse transmitted by the opposite cavity [Fig.~\ref{scheme1}(b) and \ref{scheme1}(e)]. If the atom is in $\ket{g_{2}}$, then the output single-photon pulse exits from the same cavity through which the input pulse entered [Fig.~\ref{scheme1}(c)]. Due to the finitude of the cooperativity parameter, the bright component of the input field can be absorbed by the atom-cavity system, rather than being completely reflected, and then transmitted by the wrong output port of the cavity system [blue line in Fig.~\ref{scheme1}(e)]. This possibility decreases as the cooperativity increases.

Therefore, depending on the atomic population, the system performs controlled operations on the external fields. For instance, defining the \textit{control} qubit through the atomic ground states $\ket{g_1}\to\ket{\bar{1}}_{c}$ and $\ket{g_2}\rightarrow\ket{\bar{0}}_{c}$, the system performs a CNOT gate when the \textit{target} qubit is defined through the two-mode reservoir states $\ket{1}_{\alpha} \ket{0}_{\beta}\rightarrow\ket{\bar{1}}$ and $\ket{0}_{\alpha} \ket{1}_{\beta}\rightarrow\ket{\bar{0}}$. The overline indicates a qubit value. The system can implement a Fredkin gate when considering the two target qubits as $\ket{1}_{\alpha}\rightarrow\ket{\bar{1}}_{a}$, $\ket{1}_{\beta}\rightarrow\ket{\bar{1}}_{b}$ (analogously to $\ket{\bar{0}}_{a,b}$) [Fig.~\ref{scheme1}(d)]. Here, $\ket{1}_{\alpha}$ and $\ket{1}_{\beta}$ represent the single-photon state of the input/output pulses of cavities $a$ and $b$ ($\ket{0}_{\alpha}$ and $\ket{0}_{\beta}$ are the vacuum states), respectively. Consequently, universal quantum computation~\cite{chuang} can be performed with this setup.

This paper is organized as follows. Section~\ref{sec:model} outlines the model that describes the open quantum system under consideration. In Sec.~\ref{sec:methods}, we introduce the methods employed to investigate the system response via input-output theory. We present our main results in Sec.~\ref{sec:results}, while Sec.~\ref{sec:conclusions} covers our conclusions.

\section{Model} \label{sec:model}
The setup in Fig.~\ref{scheme1}(a), in an interaction picture rotating at the cavities frequency ($\omega_c$), and under the \textit{white-noise limit} \cite{qnoise}, is described by ($\hbar = 1$)
    \begin{align}\label{Hg}
        &H = \int^{\infty}_{-\infty}\!\!\!\!d\omega \, \omega A^{\dagger}_{\omega} A_{\omega} + \int^{\infty}_{-\infty}\!\!\!\!d\omega \, \omega B^{\dagger}_{\omega} B_{\omega} \nonumber\\
        &+\dfrac{i} {\sqrt{\pi}}\int^{\infty}_{-\infty}\!\!\!\!d\omega \left[\sqrt{\kappa_{a}}(a^{\dagger}A_{\omega}-a A^{\dagger}_{\omega}) 
        + \sqrt{\kappa_{b}}(b^{\dagger}B_{\omega}-b B^{\dagger}_{\omega})\right] \nonumber\\
        &+ \underbrace{(g_{a}a+g_{b}b)\sigma_{+}^{1} + (g^{*}_{a}a^{\dagger}+g^{*}_{b}b^{\dagger})\sigma_{-}^{1}}_{H_\text{sys}},
    \end{align}
in which $\sigma_{+}^{1}=\ket{e}\bra{g_{1}}$ and $\sigma_{-}^{1}=\ket{g_{1}}\bra{e}$, $a$ and $b$ ($a^{\dagger}$ and $b^{\dagger}$) are the annihilation (creation) operators of the intracavity modes, while $A_{\omega}$ and $B_{\omega}$ ($A^{\dagger}_{\omega}$ and $B^{\dagger}_{\omega}$) are the frequency-dependent annihilation (creation) operators~\cite{Loudon2000} of the bosonic reservoirs of cavities $a$ and $b$, respectively. The coupling between external and intracavity modes is established by the decay rates of the field amplitudes of cavities $a$ ($\kappa_a$) and $b$ ($\kappa_b$). Finally, $g_a$ and $g_b$ are the coupling strength between the atomic transition $\ket{g_{1}} \leftrightarrow \ket{e}$ and the intracavity modes.  

We consider an initial state that includes all possible inputs for the two- and three-qubit operations 
\begin{align}
    \label{ISa}        \vert \Psi(t_0) \rangle&=\underbrace{\left(\lambda_{1}\ket{g_1}+\lambda_{2}\ket{g_2}\right)}_{\text{Atom}}   \underbrace{(\ket{0}_{a}   \ket{0}_{b})}_{\text{Cavities}} \nonumber\\
    & \times \underbrace{\left(\mu_{a}\ket{1}_{\alpha} \ket{0}_{\beta}+ \mu_{b}\ket{0}_{\alpha} \ket{1}_{\beta} + \mu_{c}\ket{1}_{\alpha} \ket{1}_{\beta}\right)}_{\text{Reservoirs}},
\end{align}
in which $\ket{1}_{\alpha}=\int^{\infty}_{-\infty}d\omega \xi_\text{in}(\omega)A^{\dagger}_{\omega}\ket{0}_{\alpha}$ describes the input field in a continuous-mode single-photon state~\cite{Loudon2000} impinging upon the cavity $a$, and $\ket{1}_{\beta}=\int^{\infty}_{-\infty}d\omega \zeta_\text{in}(\omega)B^{\dagger}_{\omega}\ket{0}_{\beta}$ the analog for cavity $b$. The Fourier transform of the spectral density functions, $\xi_\text{in}(\omega)$ and $\zeta_\text{in}(\omega)$, provides the square-normalized temporal shape of the incoming pulses, $\alpha_\text{in}(t)$ and $\beta_\text{in}(t)$~\cite{Loudon2000}, while the coefficients $\lambda_k$ ($k=1,2$) and $\mu_{p}$ ($p=a,b,c$) are related to the initial probability of the respective state. 

\section{Methods} \label{sec:methods}
The system response can be investigated via the input-output theory~\cite{walls2008}, with relations 
\begin{equation}
    z_\text{out}(t) = \sqrt{2\kappa_z} \, z(t) - z_\text{in}(t), \label{inoutrel}
\end{equation}
in which
\begin{align}
    z_\ell(t) =  \tfrac{(-1)^{\delta_{\ell,\text{in}}}}{\sqrt{2\pi}} \int^{\infty}_{-\infty}d\omega e^{-i \omega(t-t_\ell)} z_\ell(\omega), \label{inop}
\end{align}
with $z_\ell(\omega) = Z_\omega(t_\ell)$, for $\ell = \{\text{in, out}\}$, $\{z=a, Z=A\}$ and $\{z=b, Z=B\}$. The dynamics of an arbitrary system operator $\mathcal{O}(t)$ is governed by the Heisenberg-Langevin equation~\cite{qnoise}
    \begin{align}\label{iogen}
        \dot{\mathcal{O}} = &-i[\mathcal{O},H_\text{sys}] - \textstyle\sum\nolimits_{l=1}^{2}\Gamma_{l}([\mathcal{O},\sigma^{l}_{+}]\sigma^{l}_{-} - \sigma^{l}_{+}[\mathcal{O},\sigma^{l}_{-}]) \nonumber \\
        &- \textstyle\sum\nolimits_{z=a}^{b}[\mathcal{O},z^{\dagger}](\kappa_{z}z  -  \sqrt{2\kappa_{z}}z_\text{in}) \nonumber \\
        &+ \textstyle\sum\nolimits_{z=a}^{b}(\kappa_{z}z^\dagger  -  \sqrt{2\kappa_{z}}z_\text{in}^\dagger)[\mathcal{O},z],
    \end{align}
in which $\Gamma_l$ is the decay rate of the atomic spontaneous emission from $\ket{e}$ to $\ket{g_l}$, and $\sigma_{+}^{l} = (\sigma_{-}^{l})^\dagger = \ket{e}\bra{g_l}$. Note that the coupling of each atomic transition to its respective reservoir was also considered in Eq.~\eqref{iogen} [omitted in Eq.~\eqref{Hg}], establishing the exclusive source of photon loss considered in our setup. However, without loss of generality, we suppress the input operators for these reservoirs, as they are assumed to be initially in the vacuum state.

The set of coupled dynamical equations for the system operators based on the Heisenberg-Langevin equations is nonlinear and challenging to solve. However, it becomes linear by applying the Holstein-Primakoff approximation $\sigma_z = \sigma_{+}^{1}\sigma_{-}^{1} - \sigma_{-}^{1}\sigma_{+}^{1} \approx -\mathbf{1}$~\cite{holsteinprimakoff}, which holds as long as the atom is rarely excited and becomes exact if the atom only populates $\ket{g_2}$ or $g_a=g_b=0$. Thus,
\begin{align}
    \dot{a}(t) &= -ig_a^{*}\sigma_{-}^{1}(t) - \kappa_a a(t) + \sqrt{2\kappa_a} a_\text{in}(t) \label{adot}, \\
    \dot{b}(t) &= -ig_b^{*}\sigma_{-}^{1}(t) - \kappa_b b(t) + \sqrt{2\kappa_b} b_\text{in}(t) \label{bdot}, \\
    \dot{\sigma}_{-}^{1}(t) &= -i[g_a a(t) + g_b b(t)] - \Gamma\sigma_{-}^{1}(t), \label{sdot}
\end{align}
with $\Gamma = \Gamma_1 + \Gamma_2$. As detailed in Appendix \ref{app:analytical}, an analytical input-output relation can be derived for the field amplitudes in the frequency domain. Setting (just for convenience)  
$g_a=-g_b \equiv g \in \mathbb{R}$ and $\kappa_a = \kappa_b \equiv \kappa$, we have 
\begin{align} \label{holpri}
    a_\text{out}^{\dagger}(\omega) &= r_\omega a_\text{in}^{\dagger}(\omega) + t_\omega b_\text{in}^{\dagger}(\omega), \\
    b_\text{out}^{\dagger}(\omega) &= t_\omega a_\text{in}^{\dagger}(\omega) + r_\omega b_\text{in}^{\dagger}(\omega),
\end{align}
where $r_\omega$ and $t_\omega$ are the (frequency- and atomic-state-dependent) complex reflection and transmission coefficients, respectively. 
For resonant incoming fields, the system response is
\begin{align} 
    r_{\omega_c} &=  \frac{1}{1+4C} \, \ket{g_1}\bra{g_1} + \ket{g_2}\bra{g_2}, \label{ref} \\
    t_{\omega_c} &= \frac{4C}{1+4C} \, \ket{g_1}\bra{g_1}, \label{trans}
\end{align}
in which $C = g^2/2\kappa\Gamma$ is the cooperativity parameter. It is also worth examining the system response through the collective-mode operators $X_\text{in(out)}^{\pm}(\omega) = [a_\text{in(out)}(\omega) \pm b_\text{in(out)}(\omega)]/\sqrt{2}$, such that $X_\text{out}^{-}(\omega) = (r_{\omega} - t_{\omega}) X_\text{in}^{-}(\omega)$ and $X_\text{out}^{+}(\omega) = X_\text{in}^{+}(\omega)$.

In a regime beyond the validity of the Holstein-Primakoff approximation, it is advantageous to work with the dynamical equations for the average values of the operators. The resulting set is linear but infinite; however, it can become finite but nonlinear by applying the \textit{semiclassical} mean-field approximation~\cite{PhysRevA.2.336} (e.g., $\langle \sigma a \rangle \approx \langle \sigma \rangle \langle a \rangle$), culminating in a set of dynamical equations for $\moy{a}$, $\moy{b}$, $\moy{\sigma_{-}^{1}}$, $\moy{\sigma_{z}}$ and $\moy{\sigma_{ee}}$, with $\sigma_{ee} = \ket{e}\bra{e}$ (see Appendix \ref{app:semi}).
The solution to the semiclassical dynamical equations can be obtained numerically without difficulty, but this solution neglects the correlation between the atom and the intracavity fields, and assumes the external fields as classical ones (coherent pulses), such that $\moy{a_\text{in}}= \alpha_\text{in}(t)$ and $\moy{b_\text{in}}= \beta_\text{in}(t)$. Both analytical and semiclassical approaches will be used here to compare them with the quantum description of the setup outlined below. This analysis allowed us to investigate the influence of nonlinearities and quantum correlations on the implementation of quantum gates in our setup.

When the initial state contains only a single excitation ($\mu_c = 0$), an \textit{exact} quantum approach, excluding the occurrences of photon loss due to spontaneous atomic emission, can be obtained through the Schrödinger equation $i\partial_{t}\ket{\Psi(t)}=H_\text{eff}\ket{\Psi(t)}$, with $H_\text{eff} = H - i\Gamma \sigma_{ee}$ \cite{kuhn2012}. In this scenario, which allows describing the CNOT gate operations at least, the temporal shapes of the output pulses emerging from cavity $a$ [$\alpha^{l}_\text{out}(t)$] and $b$ [$\beta^{l}_\text{out}(t)$], conditioned to the atomic state $\ket{g_l}$, are determined using the input-output relations 
\begin{align}
    \alpha^{l}_\text{out}(t)&=\sqrt{2\kappa_{a}}c_{a}^{l}(t) - \alpha^{l}_\text{in}(t), \label{alpha} \\
    \beta^{l}_\text{out}(t)&=\sqrt{2\kappa_{b}}c_{b}^{l}(t) - \beta^{l}_\text{in}(t), \label{beta}
\end{align}
with $\alpha^{l}_\text{in}(t) = \lambda_l \mu_a \alpha_\text{in}(t)$ and $\beta^{l}_\text{in}(t) = \lambda_l \mu_b \beta_\text{in}(t)$, $c_{a}^{l}$ and $c_{b}^{l}$ being the probability amplitudes of finding a single excitation inside the cavity $a$ and $b$, respectively, while the atom is in $\ket{g_l}$ and the other modes are in the vacuum state (see Appendix \ref{app:exact}). The non-Hermitian nature of $H_\text{eff}$ results in an unnormalized state $\ket{\Psi(t)}$, such that $(1-\int_{-\infty}^{\infty}\left|\bra{\Psi(t)}\Psi(t)\rangle\right|^2 dt)$ quantifies the probability of losing the input single photon due to atomic spontaneous emission, corresponding to the failure probability of the CNOT gate. Here, we describe the input pulse as a Gaussian wave packet that contains a single photon, $\alpha_\text{in}(t) \text{ or } \beta_\text{in}(t)= (\eta\sqrt{\pi})^{-\frac{1}{2}}\exp[{-\frac{1}{2}\frac{(t-t_{0})^2}{\eta^2}}] $, in which $t_0$ is the time when its maximum reaches the cavity semi-transparent mirror and $\tau_{p}=2\eta\sqrt{2\ln(2)}$ is the pulse duration. We assume that $\tau_p$ is sufficiently long to ensure that the pulse spectral spread ($\tau_p^{-1}$) fits within the linewidth of the cavity ($2\kappa$), i.e., the entire input pulse enters and exits the empty cavity, resulting in an output pulse that maintains the shape of the input pulse~\cite{cvbpra18}.

It only remains to determine an exact solution for the system dynamics associated with the three-qubit operation with one incoming single-photon pulse for each cavity ($\mu_c =1$). However, the previous method leads to a set of integro-differential equations that becomes challenging to solve as we increase the number of excitations in the setup. To circumvent that, we opted to employ the method of hierarchical master equations~\cite{baragiola} to investigate the exact system response when the input field is in the state $\ket{1}_{\alpha} \ket{1}_{\beta}$
\begin{align}
    &\dot{\varrho}_{m,n;p,q}(t) = -i\left[H_\text{sys},\varrho_{m,n;p,q}\right]   \nonumber\\
    &+\sqrt{m}\alpha_\text{in}(t)[\varrho_{m-1,n;p,q},L^{\dagger}_{a}] + \sqrt{p}\beta_\text{in}(t)[\varrho_{m,n;p-1,q},L^{\dagger}_{b}] \nonumber\\
    &+\sqrt{n}\,\alpha_\text{in}^{*}(t)[L_{a},\varrho_{m,n-1;p,q}] + \sqrt{q}\,\beta_\text{in}^{*}(t)[L_{b},\varrho_{m,n;p,q-1}] \nonumber \\
    &+ \textstyle\sum_{z=a}^{b}\mathcal{L}[L_{z}]\varrho_{m,n;p,q}  + \textstyle\sum_{j=1}^{2}\mathcal{L}[L_{j}]\varrho_{m,n;p,q}, \label{hierarc}
\end{align}
in which $L_z = \sqrt{2\kappa_z}\,z$ and $L_j = \sqrt{2\Gamma_j}\,\sigma_{-}^{j}$, with $\mathcal{L}[L]\varrho = L^\dagger\varrho L - (L^\dagger L \varrho +\varrho L^\dagger L)/2$. For the biphoton initial state, the system dynamics is described by $\varrho_{1,1;1,1}(t)$ with the initial conditions $\varrho_{m,n;p,q}(0) = \rho_\text{sys}(0)$ if $m=n$ and $p=q$, being $\varrho_{m,n;p,q}(0) = 0$ otherwise. Here, $\rho_\text{sys}(0)$ is the density matrix of the system initial state related to the first line of Eq.~\eqref{ISa}. The mean photon fluxes of the output fields are $\moy{a_\text{out}^\dagger(t) a_\text{out}(t)} = \mathbb{E}_{1,1;1,1}[L_a^\dagger L_a] + \alpha_\text{in}^{*}(t) \mathbb{E}_{0,1;1,1}[L_a] + \mathbb{E}_{1,0;1,1}[L_a^\dagger] + |\alpha_\text{in}(t)|^2$ and $\moy{b_\text{out}^\dagger(t) b_\text{out}(t)} = \mathbb{E}_{1,1;1,1}[L_b^\dagger L_b] + \beta_\text{in}^{*}(t) \mathbb{E}_{1,1;0,1}[L_b] + \mathbb{E}_{1,1;1,0}[L_b^\dagger] + |\beta_\text{in}(t)|^2$, in which $ \mathbb{E}_{m,n;p,q}[L] = \text{Tr}_\text{sys}[\rho_{m,n;p,q}^\dagger\,L]$.

\section{Results} \label{sec:results}
Hereafter we assume resonant incoming fields, $g_a=-g_b \equiv g \in \mathbb{R}$, $\Gamma_1$ = $\Gamma_2$ and $\kappa_a = \kappa_b \equiv \kappa$, unless otherwise stated. First, note that any operation with the atom in $\ket{g_2}$ ($\ket{g_2} \ket{\psi}_\text{in} \to \ket{g_2} \ket{\psi}_\text{out}$) invariably succeeds ($r_{\omega_c} = 1$), since this state is not coupled to the cavity modes. Therefore, the system operates as two independent empty cavities, and the output pulse exits from the same cavity through which the input pulse entered. Furthermore, regardless of the atomic state, the operations involving $\ket{\psi}_\text{in} = \ket{0}_\alpha \ket{0}_\beta$ also succeed because of their trivial dynamics. 

\begin{figure}[t]
\includegraphics[trim = 0mm 9mm 0mm 8mm, width =.47\columnwidth, clip]{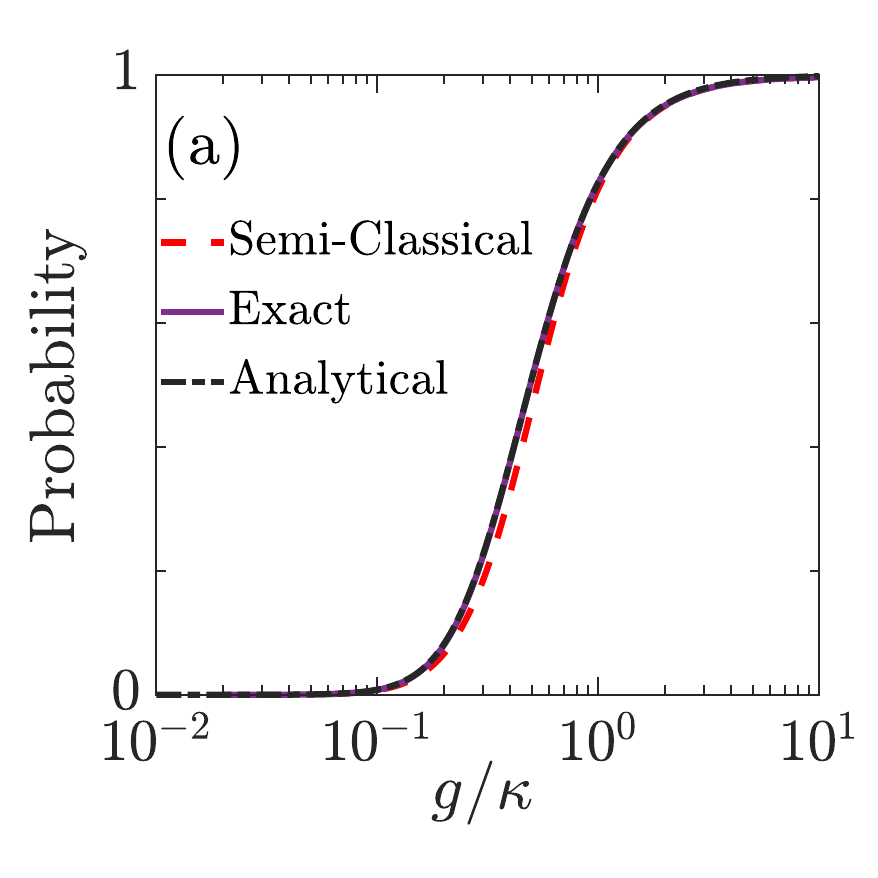}
\includegraphics[trim = 0mm 9mm 0mm 8mm, width =.47\columnwidth, clip]{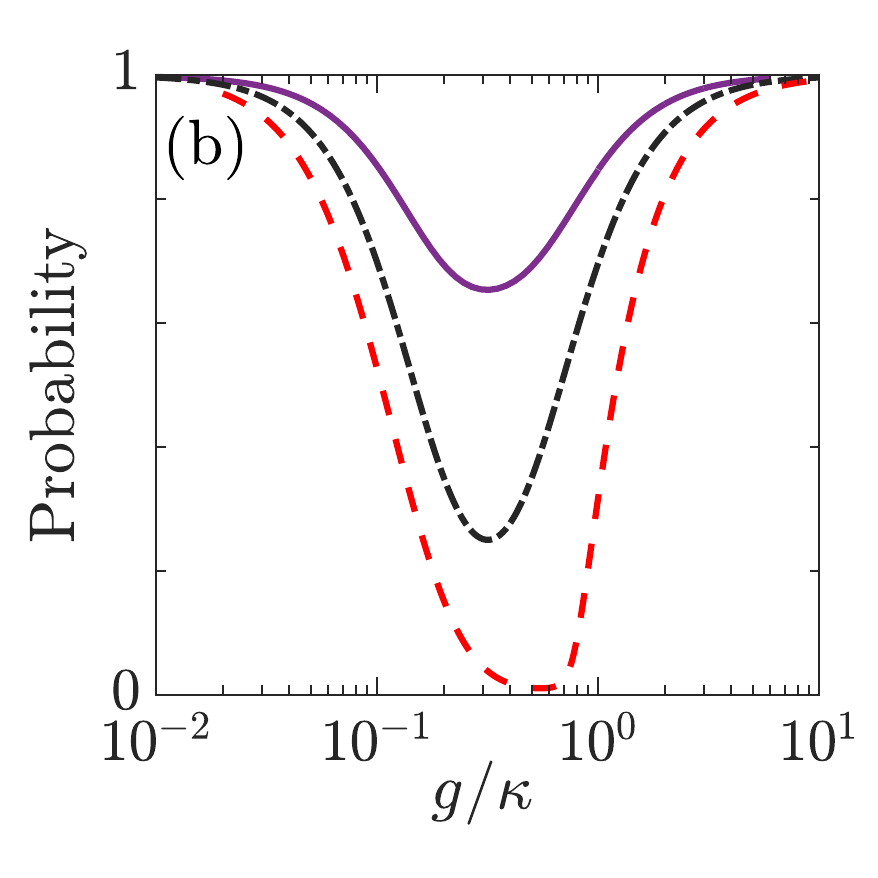}
\caption{\label{comp} Comparison among the analytical, semiclassical and exact approaches for the probabilities related to (a) $\ket{g_1} \ket{0}_\alpha \ket{1}_\beta \to \ket{g_1} \ket{1}_\alpha \ket{0}_\beta$ and (b) $\ket{g_1} \ket{1}_\alpha \ket{1}_\beta \to \ket{g_1} \ket{1}_\alpha \ket{1}_\beta$ operations, using $\Gamma = 0.2\kappa$ and $\kappa \tau_p = 40$.}
\end{figure}

Both gates proposed here contain the operations $\ket{g_1} \ket{0}_\alpha \ket{1}_\beta \to \ket{g_1} \ket{1}_\alpha \ket{0}_\beta$ and $\ket{g_1} \ket{1}_\alpha \ket{0}_\beta \to \ket{g_1} \ket{0}_\alpha \ket{1}_\beta $. From Eqs.~\eqref{ref} and \eqref{trans}, the analytical probability of their occurrence is $\vert \bra{g_1} t_{\omega_c} \ket{g_1} \vert^2 = [4C/(1+4C)]^2$. For comparison purposes, Figure~\ref{comp}(a) exhibits the analytical result alongside the semiclassical ($\int_{-\infty}^{\infty}dt \, \vert \moy{a_\text{out}(t)} \vert^2$) and exact ($\int_{-\infty}^{\infty}dt \, \vert \alpha_\text{out}^{1}(t) \vert^2$) results as a function of $g/\kappa$, considering $\Gamma = 0.2\kappa$. It is worth noting the equivalence in all treatments, which is expected since there is only one excitation throughout the process, and there are no appreciable light-matter correlations, leading to an approximate linear-optics regime for the given set of parameters. For the biphoton operation $\ket{g_1} \ket{1}_\alpha \ket{1}_\beta \to \ket{g_1} \ket{1}_\alpha \ket{1}_\beta$, we obtain the analytical probability $\vert \bra{g_1} (t_{\omega_c}^2 + r_{\omega_c}^2) \ket{g_1} \vert^2 =  [1+(4C)^2]^2/(1+4C)^4$, as shown in Fig.~\ref{comp}(b), where the exact ($\int_{-\infty}^{\infty}dt dt' \,  \moy{a_\text{out}^\dagger(t)a_\text{out}(t)b_\text{out}^\dagger(t')b_\text{out}(t')} $)~\cite{PhysRevA.57.2134} and semiclassical ($\int_{-\infty}^{\infty}dt \, \vert \moy{a_\text{out}(t)}\vert^2 \int_{-\infty}^{\infty}dt' \, \vert \moy{b_\text{out}(t')} \vert^2$) results are also plotted. We observe that all approaches also exhibit similar results for both low- and high-cooperativity regimes in this case. However, there is a discrepancy in the intermediate-cooperativity regime, highlighting the impact of nonlinearities and quantum correlations introduced by the presence of two photons in this process.

The system behavior can be described more elegantly and clearly using the dark and bright states of light, defined in terms of symmetric and antisymmetric collective-mode operators \cite{Delanty2011,Mximo2021}. For the subspace of $N$ photons, they are $\ket{\Psi_\mathcal{D}^N}_\text{in(out)} = \tfrac{\left[(X^{+}_\text{in(out)})^N\right]^{\dagger}}{\sqrt{N !}}\ket{0}_{\alpha}\ket{0}_{\beta}$ and $\ket{\Psi_\mathcal{B}^N}_\text{in(out)} = \tfrac{\left[(X^{-}_\text{in(out)})^N\right]^{\dagger}}{\sqrt{N !}}\ket{0}_{\alpha}\ket{0}_{\beta}$, respectively. The input field in a dark state does not interact with the atom regardless of its state (empty-cavity scenario), so $\ket{g_j}\ket{\Psi_\mathcal{D}^N}_\text{in} \to \ket{g_j}\ket{\Psi_\mathcal{D}^N}_\text{out}$. On the other hand, when the input field is in a bright state, although we also expect an empty-cavity scenario if the atom is in $\ket{g_2}$, in a high-cooperativity regime the output pulse acquires a phase $(-1)^N$ if the atom is in $\ket{g_1}$, that is, $\ket{g_j}\ket{\Psi_\mathcal{B}^N}_\text{in} \to (-1)^{jN}\ket{g_j}\ket{\Psi_\mathcal{B}^N}_\text{out}$. As a consequence, a CNOT gate can also be implemented with the control qubit defined through the single-photon dark and bright states of light, $\ket{\bar{0}}_{c}\equiv\ket{\Psi_\mathcal{D}^1} = \left(\ket{1}_{\alpha}\ket{0}_{\beta} + \ket{0}_{\alpha}\ket{1}_{\beta}\right)/\sqrt{2}$ and $\ket{\bar{1}}_{c}\equiv\ket{\Psi_\mathcal{B}^1} = \left(\ket{1}_{\alpha}\ket{0}_{\beta} - \ket{0}_{\alpha}\ket{1}_{\beta}\right)/\sqrt{2}$, whereas the target qubit is defined through the atomic superposition states $\ket{\bar{0}}\equiv(\ket{g_{2}}+\ket{g_{1}})/\sqrt{2}$ and $\ket{\bar{1}}\equiv(\ket{g_{2}}-\ket{g_{1}})/\sqrt{2}$.

\begin{figure}[t]
\centering
\includegraphics[trim = 8mm 7mm 15mm 9mm, width = 0.75\columnwidth, clip]{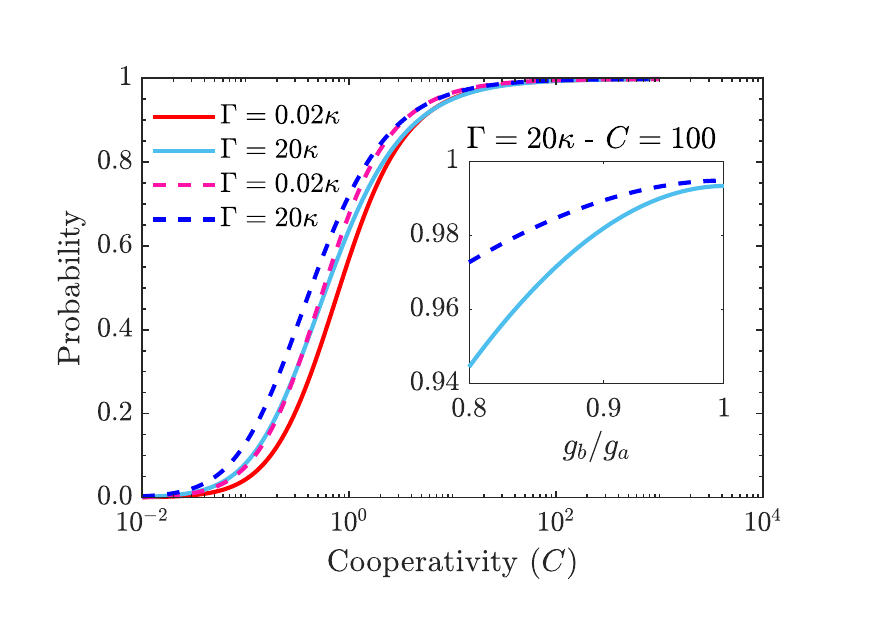}
\caption{\label{genfid} Probabilities as a function of the cooperativity $C$, for different values of total atomic spontaneous decay $\Gamma$, related to the operations $\ket{g_1} \ket{0}_\alpha \ket{1}_\beta \to \ket{g_1} \ket{1}_\alpha \ket{0}_\beta$ with the atom as the control qubit (solid lines), and $\ket{\Psi_\mathcal{B}^N}\ket{\bar{1}} \to \ket{\Psi_\mathcal{B}^N}\ket{\bar{0}}$ (or $\ket{\Psi_\mathcal{B}^N}\ket{\bar{0}} \to \ket{\Psi_\mathcal{B}^N}\ket{\bar{1}}$) with the two modes as the control qubit. In the inset we show the sensitivity of the fidelity to deviations in the values of $g_a$ compared to $g_b$. We set $\kappa\tau_{p}=40$.}
\end{figure}

Figure \ref{genfid} shows the exact probability of the occurrence of the operation $\ket{\bar{1}}_c \ket{\bar{0}} \to \ket{\bar{1}}_c \ket{\bar{1}}$ (identical result for $\ket{\bar{1}}_c \ket{\bar{1}} \to \ket{\bar{1}}_c \ket{\bar{0}}$) when considering the atom controlling the light states (solid lines) and vice versa (dashed lines), for low ($\Gamma = 0.02\kappa$) and high ($\Gamma = 20\kappa$) atomic spontaneous emission rates. This analysis summarizes the probability of success in implementing the CNOT gate, since only these operations do not have unitary fidelity for all ranges of parameters in this case. We observe a probability greater than 95\% for $C \gtrsim 10$. Remarkably, this high probability occurs not only for the strong-coupling regime ($\Gamma = 20\kappa \text{ and } C \sim 10 \Rightarrow g=20\kappa$), where the reflection of the input pulse in a bright state is expected when the atom is in $\ket{g_1}$ due to the Autler-Townes splitting~\cite{autlertownes}, but also for the weak-coupling regime ($\Gamma = 0.02\kappa \text{ and } C \sim 10 \Rightarrow g\approx0.6\kappa$), where the reflection is due to a destructive quantum interference between the system absorption paths (electromagnetically-induced-transparency-like phenomenon) \cite{fleisch,Alsing,Rice1996}. The collective effects of light used in our protocol yield a failure probability 50\% lower than that of the Duan-Kimble scheme \cite{duan04} (see Appendix~\ref{app:comparison}), which considers a single cavity. Furthermore, although there is a success probability of 95\% for this operation when $C \simeq 10$, it can reach a post-selected fidelity of 99.9\% for the same cooperativity value after detecting a photon emerging from the cavity system (see Appendix~\ref{app:post}). Furthermore, the inset of Fig.~\ref{genfid} shows a small deviation in the success probability of performing a CNOT gate as the value of $g_a$ deviates from the value of $g_b$ (similar for deviations between $\kappa_a$ and $\kappa_b$). This deviation changes the form of the bright and dark states; then the quantum interference is no longer totally achieved, decreasing the success probability.

\begin{figure}[tb]
\includegraphics[trim = 0.7cm 0.35cm 1.55cm 1.1cm, width =0.49\columnwidth, clip]{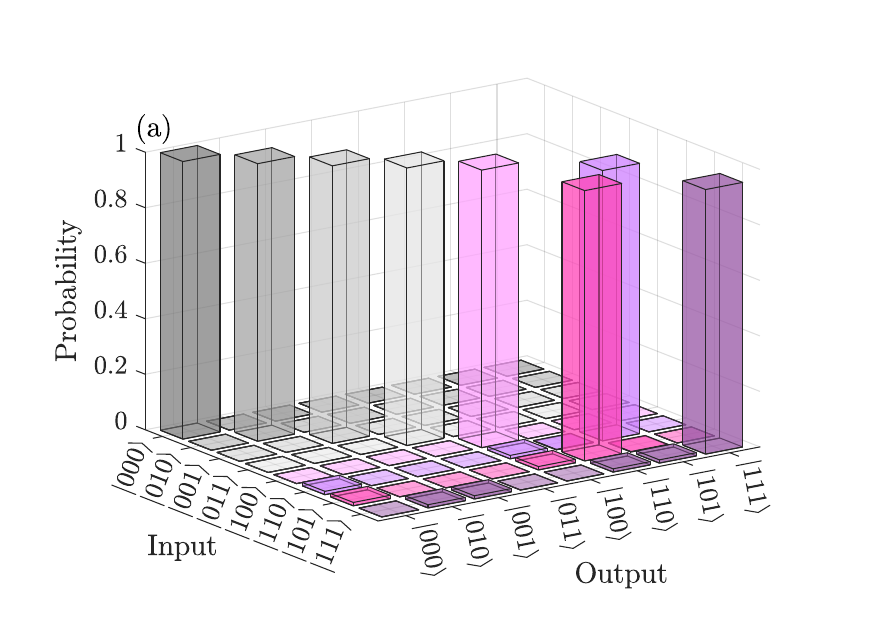}
\includegraphics[trim = 0.7cm 0.35cm 1.55cm 1.1cm, width =0.49\columnwidth,clip]{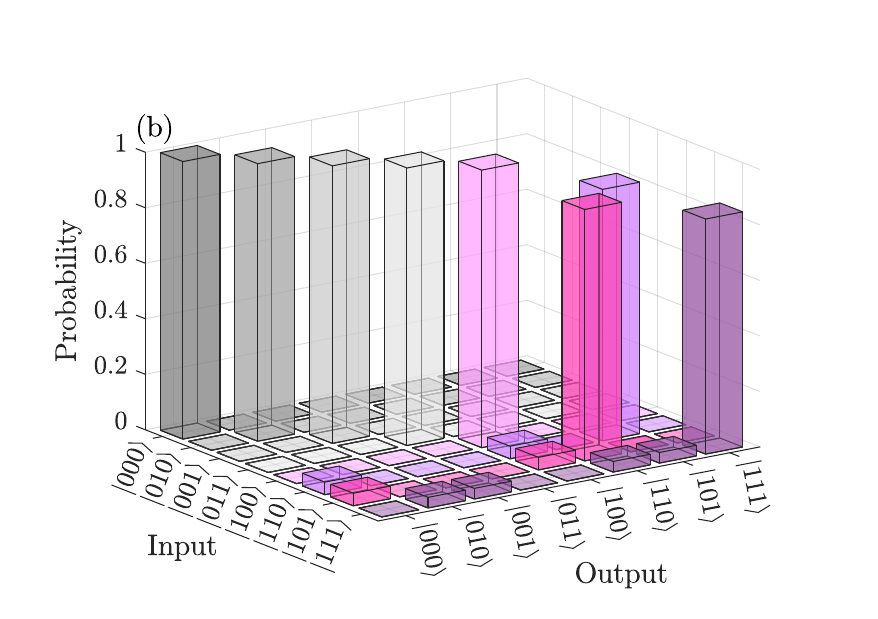}
\caption{\label{bar3d}Probabilities of the Fredkin gate operations for (a) $C=20$ and (b) $C=5$, considering $\Gamma=0.2\kappa$ and $\kappa \tau_p = 40$.}
\end{figure}

Finally, Figure~\ref{bar3d} illustrates the probability of occurrence for all operations in the implementation of a Fredkin gate [truth table in Fig.~\ref{scheme1}(d)], considering $\kappa \tau_p = 40$ and $\Gamma = 0.2\kappa$ for two experimentally feasible values of cooperativity~\cite{brekenfeld}, $C = 5$ [Fig.~\ref{bar3d}(b)] and $C=20$ [Fig.~\ref{bar3d}(a)]. We observe a success probability greater than 95\% for $C=20$. However, we can notice a decrease of that for $C=5$ (probability of 85\%), where the losses by atomic emission become evident, mostly for the operation involving two photons.

\section{Conclusions} \label{sec:conclusions}
We have shown that a cross-cavity system coupled to a $\Lambda$-type three-level atom provides a promising platform to perform universal quantum computation. We demonstrated how to implement a CNOT gate based on a $\pi$-phase gate that involves bright and dark states of light, which is activated depending on the atomic state. We can have either the atom controlling the light states or vice versa. 
Remarkably, we have shown that a quantum Fredkin gate can also be directly implemented with this system, where the atom controls the light states of both pulses interacting with the cavities. Based on the state-of-the-art parameters~\cite{brekenfeld}, our results predict high success probabilities for the two- and three-qubit quantum gates investigated in this study. Our system can also be extended to the case of multiple pulses or multiple atoms within the crossed-cavity system, thus allowing the implementation of multi-qubit logic gates~\cite{spread2}. 
Therefore, our results contribute to the advancement of quantum technologies for applications in quantum information, communication, and computation.

\begin{acknowledgments}
We acknowledge discussions in the earlier stage of this work with Prof. Dr. Gerhard Rempe and the fruitful discussions with Prof. Dr. Antonio S. M. de Castro. This work was supported by the S\~ao Paulo Research Foundation (FAPESP, Grants Nos. 2019/11999-5, 2018/22402-7, and 2022/00209-6) and by the Brazilian National Council for Scientific and Technological Development (CNPq, Grants Nos. 405712/2023-5, 311612/2021-0 and 465469/2014-0). 

\end{acknowledgments}

\appendix

\section{Heisenberg-Langevin equations with Holstein-Primakoff approximation -- analytical method} \label{app:analytical}
The Fourier transform of Eqs.~(\ref{adot})--(\ref{sdot}) yields
    \begin{align}\label{h-l ft}
        &-i\omega a(\omega) = -ig_{a}^{*}\sigma^{1}_{-}(\omega) - \kappa_{a} a(\omega) + \sqrt{2\kappa_{a}}a_\text{in}(\omega), \\ \label{h-l ft2}
        &-i\omega b(\omega) = -ig_{b}^{*}\sigma^{1}_{-}(\omega) - \kappa_{b} b(\omega) + \sqrt{2\kappa_{b}}b_\text{in}(\omega), \\ \label{h-l ft3}
        &-i \omega \sigma^{1}_{-}(\omega) = -i[g_{a}a(\omega) + g_{b}b(\omega)] - \Gamma\sigma_{-}^{1}(t).
    \end{align}
Considering just for convenience $g_{a} = -g_{b} \equiv g \in \mathbb{R}$ and $\kappa_a = \kappa_b \equiv \kappa$, we have from Eq.~(\ref{h-l ft3}) that
    \begin{align}
        \sigma_{-}^{1}(\omega) &= -\frac{i g}{\Gamma - i\omega}(a(\omega) - b(\omega)) \nonumber \\ 
        &= -\frac{i g \sqrt{2}}{(\Gamma - i\omega)}X^{-}(\omega), \label{sm}
    \end{align}
where the collective-mode operators are defined as
    \begin{align}\label{xdef}
        X^{\pm}(\omega) = \frac{1}{\sqrt{2}}[a(\omega)\pm b(\omega)].
    \end{align}
We must be cautious when expressing the atomic operator in terms of the collective-mode operator [Eq.~\eqref{sm}] to be inserted into Eqs.~\eqref{h-l ft} and \eqref{h-l ft2}, since there is a risk of overlooking a proper description of the information regarding the atom. Specifically, by performing this substitution, we disregard the fact that $\sigma_{-}^{1}(\omega) = 0$ in Eq.~\eqref{sm} when the atom is initially in $\ket{g_2}$ (which is equivalent to $g=0$). For our purposes, this mistake can be avoided by considering in an ad hoc manner the substitution $g \to \Tilde{g} = g\ket{g_1}\bra{g_1}$ on the right side of Eq.~\eqref{sm}.

Inserting Eq.~\eqref{sm} into Eqs.~\eqref{h-l ft} and \eqref{h-l ft2},
    \begin{align}\label{1}
        X^{-}(\omega) &= \frac{\sqrt{2\kappa}(\Gamma - i\omega)}{(\kappa - i\omega)(\Gamma - i\omega) + (\Tilde{g}\sqrt{2})^{2}} X_\text{in}^{-}(\omega), \\
        X^{+}(\omega) &= \frac{\sqrt{2\kappa}}{\kappa - i\omega} X^\text{in}_{+}(\omega),
    \end{align}
where we can observe that the symmetric collective-mode operator are decoupled from the atom regardless its state, since it does not depend on $g$.
From the input-output relation written in terms of the collective-mode operators,
\begin{equation}
    X_\text{out}^{\pm}(\omega) = \sqrt{2\kappa}X^{\pm}(\omega) - X_\text{in}^{\pm}(\omega),
\end{equation} 
we obtain
\begin{align}\label{eqax}
        &X_\text{out}^{-}(\omega) = x_{-}^{*} X_\text{in}^{-}(\omega),  \\ \label{eqax2}
        &X_\text{out}^{+}(\omega) = x_{+}^{*} X_\text{in}^{+}(\omega),   
    \end{align}
with
    \begin{align}
        &x_{-}^{*} \equiv \frac{(\kappa + i\omega)(\Gamma - i\omega) - (\Tilde{g}\sqrt{2})^{2}}{(\kappa - i\omega)(\Gamma- i\omega) + (\Tilde{g}\sqrt{2})^{2}},\\
        &x_{+}^{*} \equiv \left(\frac{\kappa + i\omega}{\kappa - i\omega}\right),
    \end{align}
directly implying in
    \begin{align}
        &a_\text{out}^{\dagger}(\omega) = \underbrace{\tfrac{(x_{+} + x_{-})}{2}}_{r_\omega} a_\text{in}^{\dagger}(\omega) + \underbrace{\tfrac{(x_{+} - x_{-})}{2}}_{t_{\omega}}b_\text{in}^{\dagger}(\omega),\\
        &b_\text{out}^{\dagger}(\omega) = \underbrace{\tfrac{(x_{+} - x_{-})}{2}}_{t_{\omega}} a_\text{in}^{\dagger}(\omega) + \underbrace{\tfrac{(x_{+} + x_{-})}{2}}_{r_{\omega}}b_\text{in}^{\dagger}(\omega),
    \end{align}
with $r_\omega$ and $t_\omega$ being the (frequency- and atomic-state-dependent) complex reflection and transmission coefficients.

At this point it is important remarking that the Hamiltonian in Eq.~\eqref{Hg} was written in an interaction picture rotating at the cavities frequency. The results for the Heisenberg picture are obtained by making $\omega \to \omega - \omega_c$ in Eqs.~\eqref{eqax} and \eqref{eqax2}. Therefore, the system response for resonant incoming fields ($\omega = \omega_c$) is given by Eqs.~\eqref{ref} and \eqref{trans}.

\section{Mean-field approximation -- semiclassical method} \label{app:semi}
When the atom can be significantly excited, in a parameter regime beyond the validity of the Holstein-Primakoff approximation, it is convenient to work with the dynamical equations for the expectation values of the operators in Eqs.~\eqref{adot}--\eqref{sdot}
\begin{align}
    \dot{\moy{a}} &= -ig_a^{*}\moy{\sigma_{-}^1} - \kappa_a \moy{a} + \sqrt{2\kappa_a} \moy{a_\text{in}},  \label{adotsc}\\
    \dot{\moy{b}} &= -ig_b^{*}\moy{\sigma_{-}^1} - \kappa_b \moy{b} + \sqrt{2\kappa_b} \moy{b_\text{in}},  \label{bdotsc}\\
    \dot{\moy{\sigma_{-}^1}} &= i(g_a \moy{a\sigma_z} + g_b \moy{b\sigma_z}) - \Gamma \moy{\sigma_{-}^1}. \label{sdotsc}
\end{align}
This set of differential equations is not closed. Specifically, the equation of motion for $\moy{\sigma_{-}^1}$ depends on the expectation values $\moy{a\sigma_z}$ and $\moy{b\sigma_z}$, which involve two operators. Similarly, the equations of motion for expectation values involving two operators rely on expectation values with three operators, and so forth [known as the Bogoliubov-Born-Green-Kirkwood-Yvon (BBGKY) hierarchy]. An approach to truncate this set of equations, beyond the Holstein-Primakoff approximation, is through the semiclassical mean-field approximation~\cite{PhysRevA.2.336}, which relies on the factorization of expectation values that involve two operators, e.g., $\langle \sigma a \rangle \approx \langle \sigma \rangle \langle a \rangle$. The resulting set of dynamical equations becomes finite but nonlinear
\begin{align}
    \dot{\moy{a}} &= -ig_a^{*}\moy{\sigma_{-}^1} - \kappa_a \moy{a} + \sqrt{2\kappa_a} \moy{a_\text{in}},  \label{adotsc2}\\
    \dot{\moy{b}} &= -ig_b^{*}\moy{\sigma_{-}^1} - \kappa_b \moy{b} + \sqrt{2\kappa_b} \moy{b_\text{in}},  \label{bdotsc2}\\
    \dot{\moy{\sigma_{-}^1}} &= i(g_a \moy{a} + g_b \moy{b})\moy{\sigma_z} - \Gamma \moy{\sigma_{-}^1}, \label{sdotsc2} \\
    \dot{\moy{\sigma_{z}}} &= -2i(g_a \moy{a} + g_b\moy{b})\moy{\sigma_{-}^1}^{*} \nonumber \\ &+ 2i(g_a^{*} \moy{a}^{*} + g_b^{*}\moy{b}^{*})\moy{\sigma_{-}^1}  - 2(\Gamma_1 + \Gamma) \moy{\sigma_{ee}}, \label{szdotsc2} \\
    \dot{\moy{\sigma_{ee}}} &= -i(g_a \moy{a} + g_b\moy{b})\moy{\sigma_{-}^1}^{*} \nonumber \\ &+ i(g_a^{*} \moy{a}^{*} + g_b^{*}\moy{b}^{*})\moy{\sigma_{-}^1} - 2\Gamma \moy{\sigma_{ee}}. \label{seedotsc2}
\end{align}

\section{Schrödinger equation for single-excitation initial state -- exact method} \label{app:exact}
When the initial state of the setup contains a single excitation [$\mu_c = 0$ in Eq.~\eqref{ISa}] an exact solution can be obtained by numerically solving the Schrödinger equation $i\partial_{t}\ket{\Psi(t)}=H_\text{eff}\ket{\Psi(t)}$, with $H_\text{eff} = H - i\Gamma \sigma_{ee}$ \cite{kuhn2012}, in which the non-Hermitian damping term excludes the occurrences of photon loss due to spontaneous atomic emission. The single excitation condition allows one to consider the following general evolved state
     \begin{align}
         \label{ESa}
            \ket{\Psi&(t)} = c_{e}(t)\ket{e}\ket{0}_{a}\ket{0}_{b}\ket{0}_{\alpha}\ket{0}_{\beta}   \nonumber\\
            &+ \sum^{2}_{l=1}{c^{l}_{a}(t)\ket{g_{l}}\ket{1}_{a}\ket{0}_{b}\ket{0}_{\alpha}\ket{0}_{\beta}}    \nonumber\\
            &+ \sum^{2}_{l=1}{c^{l}_{b}(t)\ket{g_{l}}\ket{0}_{a}\ket{1}_{b}\ket{0}_{\alpha}\ket{0}_{\beta}}    \nonumber\\
            &+ \sum^{2}_{l=1}{\int^{\infty}_{-\infty}d\omega\xi_{l}(\omega,t)\ket{g_{l}}\ket{0}_{a}\ket{0}_{b}A^{\dagger}(\omega)\ket{0}_{\alpha}\ket{0}_{\beta}}  \nonumber\\
            &+ \sum^{2}_{l=1}{\int^{\infty}_{-\infty}d\omega\zeta_{l}(\omega,t)\ket{g_{l}}\ket{0}_{a}\ket{0}_{b}\ket{0}_{\alpha}B^{\dagger}(\omega)\ket{0}_{\beta}}.
     \end{align}
Inserting the evolved state into the Schrödinger equation yields the following sets of coupled integro-differential equations for the amplitudes
    \begin{align}
    \label{ideq}
        \begin{bmatrix}
            \dot{c}_{e} \\
            \dot{c}_{a}^{1} \\
            \dot{c}_{b}^{1} \\
            \dot{\xi}_{1} \\
            \dot{\zeta}_{1} 
        \end{bmatrix}
        =
        \begin{bmatrix}
            -\Gamma &   -ig_{a} & -ig_{b}   &   0   &   0   \\
            -ig^{*}_{a} &   0   &   0   &   \sqrt{\tfrac{\kappa_{a}}{\pi}}\int d\omega    &   0   \\
            0   &   -ig^{*}_{b} &   0   &   0   &   \sqrt{\tfrac{\kappa_{b}}{\pi}}\int d\omega \\
            0   &   -\sqrt{\tfrac{\kappa_{a}}{\pi}}   &   0   &   -i\omega &   0   \\
            0   &   0   &   -\sqrt{\tfrac{\kappa_{b}}{\pi}}  &   0   &   -i\omega
        \end{bmatrix}
        \begin{bmatrix}
            {c}_{e} \\
            {c}_{a}^{1} \\
            {c}_{b}^{1} \\
            {\xi}_{1} \\
            {\zeta}_{1}
        \end{bmatrix},
        \end{align}
        \begin{align}
        \begin{bmatrix}
            \dot{c}_{a}^{2} \\
            \dot{\xi}_{2}
        \end{bmatrix}
        =
        \begin{bmatrix}
            0   &   \sqrt{\kappa_{a}/\pi}\int d\omega   \\
            \sqrt{\kappa_{a}/\pi}   &   -i\omega
        \end{bmatrix}
        \begin{bmatrix}
            {c}_{a}^{2} \\
            {\xi}_{2}
        \end{bmatrix}, \\
        \nonumber \\
        \begin{bmatrix}
            \dot{c}_{b}^{2} \\
            \dot{\zeta}_{2}
        \end{bmatrix}
        =
        \begin{bmatrix}
            0   &   \sqrt{\kappa_{b}/\pi}\int d\omega   \\
            \sqrt{\kappa_{b}/\pi}   &   -i\omega
        \end{bmatrix}
        \begin{bmatrix}
            {c}_{b}^{2} \\
            {\zeta}_{2}
        \end{bmatrix}.
    \end{align}

By integrating the equations for $\dot{\xi}_{l}$ and $\dot{\zeta}_{l}$ from an initial time $t_{0}$ to $t>t_0$, we obtain
    \begin{align}
        \xi_{l}(\omega,t) &= \xi_{l}(\omega,t_0) e^{-i\omega(t - t_0)} - \sqrt{\frac{\kappa_{a}}{\pi}}\int^{t}_{t_0} \!\!\! d\tau c^{l}_{a}(\tau) e^{-i\omega(t - \tau)}, \\
        \zeta_{l}(\omega,t) &= \zeta_{l}(\omega,t_0) e^{-i\omega(t - t_0)} - \sqrt{\frac{\kappa_{b}}{\pi}}\int^{t}_{t_0} \!\!\! d\tau c^{l}_{b}(\tau) e^{-i\omega(t - \tau)},
    \end{align}
for $l=1,2$. Now we apply the time limit for these solutions, for a past time $t_0\to-\infty$ we obtain $\xi_{l}(\omega,t_{0})\equiv\xi_\text{in}^{l}(\omega)$ and $\zeta_{l}(\omega,t_{0}) \equiv \zeta_\text{in}^{l}(\omega)$ when the incoming pulse is still at a sufficiently large distance from the cavity. Also, we integrate for $t$ to a future time $t_1 > t$, such that
    \begin{align}\label{xizeta}
        \xi_{l}(\omega,t) &= \xi_{l}(\omega,t_1) e^{-i\omega(t - t_1)} + \sqrt{\frac{\kappa_{a}}{\pi}}\int^{t_{1}}_{t} \!\!\!\! d\tau c^{l}_{a}(\tau) e^{-i\omega(t - \tau)}, \\
        \zeta_{l}(\omega,t) &= \zeta_{l}(\omega,t_1) e^{-i\omega(t - t_1)} + \sqrt{\frac{\kappa_{b}}{\pi}}\int^{t_{1}}_{t} \!\!\!\! d\tau c^{l}_{b}(\tau) e^{-i\omega(t - \tau)}.
    \end{align}
Similarly for a future time $t_1\to+\infty$, we obtain $\xi_{l}(\omega,t_{1})\equiv -\xi_\text{out}^{l}$ and $\zeta_{l}(\omega,t_{1}) \equiv -\zeta_\text{out}^{l}$, where the minus sign comes from the convention that takes into account the propagation direction of the fields in their amplitudes.

The square-normalized temporal shapes of the incoming and outgoing pulses related to cavity $a$ are
    \begin{align}
    \label{ftransf}
        &\alpha_\text{in}^{l}(t) = \frac{1}{\sqrt{2\pi}}\int^{+\infty}_{-\infty}\xi_\text{in}(\omega)e^{-i\omega(t-t_0)},\\ \label{ftransf2}
        &\alpha_\text{out}^{l}(t) = \frac{1}{\sqrt{2\pi}}\int^{+\infty}_{-\infty}\xi_\text{out}(\omega)e^{-i\omega(t-t_1)},
    \end{align}
and similar for $\beta_\text{in,out}^{l}(t)$ with $\zeta^{l}_\text{in,out}$. As a result, combining Eqs.~\eqref{ftransf} and \eqref{ftransf2} with the relations obtained for $\xi_{l}(\omega,t)$ and $\zeta_{l}(\omega,t)$, we obtain Eqs.~\eqref{alpha} and \eqref{beta}, which are the {boundary conditions} relating the field amplitudes outside the cavities to the intracavity fields.

The last step to build a solvable set of equations is to eliminate the integral terms in $\dot{c}_{a}^{l}(t)$ and $\dot{c}_{b}^{l}(t)$. Let us take as an example the equation for $\dot{c}_{a}^{l}(t)$,
    \begin{align}
        \dot{c}_{a}^{l}(t) &= -ig^{*}_{a}\delta_{l,1}c_{e} + \sqrt{\frac{\kappa_{a}}{\pi}}\int_{-\infty}^{+\infty} d\omega\xi_{l}(\omega,t)\nonumber \nonumber  \\
        &= -i g_{a}^{*}\delta_{l,1} c_{e} + \sqrt{\frac{\kappa_{a}}{\pi}}\underbrace{\int^{+\infty}_{-\infty} d\omega \xi^{l}_\text{in}(\omega)e^{-i\omega(t - t_0)}}_{\sqrt{2\pi} \alpha_\text{in}^{l}(t)} \nonumber \\ & - \frac{\kappa_{a}}{\pi}\underbrace{\int^{t}_{t_0}d\tau c_{a}^{l}(\tau)\underbrace{\int^{+\infty}_{-\infty}d\omega e^{-\omega(t - \tau)}}_{2\pi \delta(t-\tau)}}_{\pi c_{a}^{l}(t)},
    \end{align}
such that
    \begin{align}
        \dot{c}_{a}^{l}(t) = -ig_{a}^{*}\delta_{l,1}c_{e}  + \sqrt{2\kappa_{a}}\alpha_\text{in}^{l}(t) - \kappa_{a}c^{l}_{a}(t).
    \end{align}
The same procedure can be taken for $c_{b}^{l}(t)$, yielding the following set of coupled differential equations
    \begin{align}
        &\begin{cases}
        \dot{c}_{a}^{1}(t)&=-ig_{a}^{*}c_{e}(t) - \kappa_{a}c_{a}^{1}(t) + \sqrt{2\kappa_{a}}\alpha_\text{in}^{1}(t) \\
        \dot{c}_{b}^{1}(t)&=-ig_{b}^{*}c_{e}(t) - \kappa_{b}c_{b}^{1}(t) + \sqrt{2\kappa_{b}}\beta_\text{in}^{1}(t) \\
        \dot{c}_{e}(t) &= -\Gamma c_{e}(t) - ig_{a}c_{a}^{1}(t) - i g_{b}c_{b}^{1}(t) 
        \end{cases}, \\
        &\begin{cases}
        \dot{c}_{a}^{2} &= -\kappa_{a}c^{2}_{a}(t) + \sqrt{2\kappa_{a}}\alpha_\text{in}^{2}(t) \\
        \dot{c}_{b}^{2} &= -\kappa_{b}c^{2}_{b}(t) + \sqrt{2\kappa_{b}}\beta_\text{in}^{2}(t)    
        \end{cases}.
    \end{align} 
Given a certain input pulse [$\alpha_\text{in} (t)$ and/or $\beta_\text{in} (t)$] and consider the initial state of Eq. (\ref{ISa}), we have that $\alpha^{l}_\text{in}(t) = \lambda_l \mu_a \alpha_\text{in}(t)$, $\beta^{l}_\text{in}(t) = \lambda_l \mu_b \beta_\text{in}(t)$ and $c_{a}^{l}(0) = c_{b}^{l}(0) = c_e(0) = 0$. Then, it is straightforward to numerically solve the system dynamics, and hence access the outgoing pulse dynamics through $\alpha^{l}_\text{out}(t)$ and $\beta^{l}_\text{out}(t)$ determined by Eqs.~\eqref{alpha} and \eqref{beta}.

\section{Comparison between the Duan-Kimble protocol and our cross-cavity scheme} \label{app:comparison}
Let us compare our proposal with the previously used atom-cavity system, which we will consider as the Duan-Kimble (DK) scheme~\cite{duan04}. The DK scheme is equivalent to considering only the cavity $a$ in Eq.~\ref{Hg}. As a result, considering the Holstein-Primakoff approximation for the sake of simplicity, one obtains, for a resonant incoming field and the atom in $\ket{g_1}$,
        \begin{equation*}
            a_\text{out} = \left(\frac{1-2C}{1+2C} \right)a_\text{in},
        \end{equation*}
        in which $C = g^2/2\kappa\Gamma$ is the ``individual'' cooperativity parameter between the atom and a single mode. We observe that a $\pi$ phase shift in this case is possible only for $C>1/2$. Moreover, the probability of not having a reflected field with a $\pi$ phase shift (failure probability) is, for $C>1/2$,
        \begin{equation}
            P_F^\text{DK} = 1-\left(\frac{1-2C}{1+2C} \right)^2 = \frac{8C}{(1+2C)^2}.
        \end{equation}
        
In our protocol, the phase shift can be acquired by the bright component of the two-mode incoming field. Therefore, we have to look at its respective collective operator
        \begin{equation*}
            X^{-}_\text{out} = \left(\frac{1-4C}{1+4C} \right)X^{-}_\text{in}.
        \end{equation*}
        In this case, a $\pi$ phase shift occurs for $C>1/4$ (minimum value corresponds to the half value for the single cavity case, which represents a gain of a factor 2 in the ``individual'' cooperativity parameter to implement a $\pi$ phase shift). Furthermore, the probability of failure in our case is, for $C>1/4$,
        \begin{equation}
            P_F = 1-\left(\frac{1-4C}{1+4C} \right)^2 = \frac{16C}{(1+4C)^2}.
        \end{equation} 
        For $C\gg1$, $\tfrac{P_F}{P_F^\text{DF}} \approx \tfrac{1}{2}$, that is, compared to the DK scheme, the probability of failure is 50\% lower in our case. For example, considering $C \approx 10$, $P_F \approx 9.52\%$ while $P_F^\text{DK}\approx 18.14\%$. This gain is due to the collective effects present in our protocol, namely the effective coupling between the atom and the bright mode of the cavity system is $g_\text{eff} = g\sqrt{2}$, then we have $C_\text{eff} = 2C$, which explains the following correspondence $a_\text{out} = \left(\tfrac{1-2C}{1+2C} \right)a_\text{in} \to X^{-}_\text{out} = \left(\tfrac{1-2C_\text{eff}}{1+2C_\text{eff}} \right)X^{-}_\text{in} =  \left(\frac{1-4C}{1+4C} \right)X^{-}_\text{in}$.

\section{Post-selected process fidelity} \label{app:post}
The implementation of a quantum gate can be characterized by two quantities: gate efficiency (success probability) and gate fidelity (accuracy of the quantum operations performed by the gate). The efficiency of gates based on cavity-assisted photon scattering~\cite{rempe2015} is affected by all sources of photon loss, including atomic spontaneous emission, absorption and scattering by beam splitters or cavity mirrors, imperfection in the photon source, and inefficiencies in photon collection and detection. However, these negative occurrences can be eliminated by post-selecting events in which a photon emitted by the cavity system is detected. Although a photon count may not be mandatory for implementing the gate, it heralds that the protocol has succeeded without modifying the gate fidelity. In other words, a photon detection just ensures us that the gate has succeeded.

In our scheme, since we have two single-sided cavities, we have two ports. When the atom is in $\ket{g_1}$ and we send a single-photon pulse towards the cavity $a$, for example, in our protocol it is desired that the photon is completely transmitted by the cavity $b$. However, it is possible to occur two other undesired events, namely the photon can be either lost (due to any source of photon loss) or come out of cavity $a$ (the photon can be absorbed by the atom-cavity system and then transmitted by the wrong output port of the cavity system). Although the first event only affects the gate probability and can be circumvented by detecting a photon emerging from the cavity system (post-selection approach), the second one corresponds to an error in the qubit value involved in the gate implementation, thus reducing the gate fidelity. 


After detecting a photon emerging from the cavity system, the post-selected fidelity of these operations is given by a quantum conditional probability~\cite{GuerraBobo2013}. 
%
%
%
%
In the implementation of the CNOT gate, only the operation $\ket{g_1}\ket{1}_\alpha\ket{0}_\beta \to \ket{g_1}\ket{0}_\alpha\ket{1}_\beta$ (or $\ket{g_1}\ket{0}_\alpha\ket{1}_\beta \to \ket{g_1}\ket{1}_\alpha\ket{0}_\beta$) has its efficiency affected, as characterized in Fig.~\ref{comp}(a). 
Thus, the post-selected fidelity of this operation after detecting a photon emerging from the cavity system is given by
\begin{equation}
    P_\text{PS} = \frac{\vert\bra{g_1}t_{\omega_c}\ket{g_1}\vert^2}{\vert\bra{g_1}t_{\omega_c}\ket{g_1}\vert^2 + \vert\bra{g_1}r_{\omega_c}\ket{g_1}\vert^2} = \frac{(4C)^2}{1+(4C)^2}.
\end{equation}
For $C \sim 10$, we have seen that this operation occurs with a probability ($P = [4C/(1+4C)]^2$) greater than 95.2\%. However, after detecting a photon emerging from the cavity system, the post-selected fidelity of this operation becomes 99.9\%.

\bibliography{apssamp}

\begin{thebibliography}{58}%
\makeatletter
\providecommand \@ifxundefined [1]{%
 \@ifx{#1\undefined}
}%
\providecommand \@ifnum [1]{%
 \ifnum #1\expandafter \@firstoftwo
 \else \expandafter \@secondoftwo
 \fi
}%
\providecommand \@ifx [1]{%
 \ifx #1\expandafter \@firstoftwo
 \else \expandafter \@secondoftwo
 \fi
}%
\providecommand \natexlab [1]{#1}%
\providecommand \enquote  [1]{``#1''}%
\providecommand \bibnamefont  [1]{#1}%
\providecommand \bibfnamefont [1]{#1}%
\providecommand \citenamefont [1]{#1}%
\providecommand \href@noop [0]{\@secondoftwo}%
\providecommand \href [0]{\begingroup \@sanitize@url \@href}%
\providecommand \@href[1]{\@@startlink{#1}\@@href}%
\providecommand \@@href[1]{\endgroup#1\@@endlink}%
\providecommand \@sanitize@url [0]{\catcode `\\12\catcode `\$12\catcode
  `\&12\catcode `\#12\catcode `\^12\catcode `\_12\catcode `\%12\relax}%
\providecommand \@@startlink[1]{}%
\providecommand \@@endlink[0]{}%
\providecommand \url  [0]{\begingroup\@sanitize@url \@url }%
\providecommand \@url [1]{\endgroup\@href {#1}{\urlprefix }}%
\providecommand \urlprefix  [0]{URL }%
\providecommand \Eprint [0]{\href }%
\providecommand \doibase [0]{https://doi.org/}%
\providecommand \selectlanguage [0]{\@gobble}%
\providecommand \bibinfo  [0]{\@secondoftwo}%
\providecommand \bibfield  [0]{\@secondoftwo}%
\providecommand \translation [1]{[#1]}%
\providecommand \BibitemOpen [0]{}%
\providecommand \bibitemStop [0]{}%
\providecommand \bibitemNoStop [0]{.\EOS\space}%
\providecommand \EOS [0]{\spacefactor3000\relax}%
\providecommand \BibitemShut  [1]{\csname bibitem#1\endcsname}%
\let\auto@bib@innerbib\@empty
\bibitem [{\citenamefont {Ladd}\ \emph {et~al.}(2010)\citenamefont {Ladd},
  \citenamefont {Jelezko}, \citenamefont {Laflamme}, \citenamefont {Nakamura},
  \citenamefont {Monroe},\ and\ \citenamefont {O'Brien}}]{Ladd2010}%
  \BibitemOpen
  \bibfield  {author} {\bibinfo {author} {\bibfnamefont {T.~D.}\ \bibnamefont
  {Ladd}}, \bibinfo {author} {\bibfnamefont {F.}~\bibnamefont {Jelezko}},
  \bibinfo {author} {\bibfnamefont {R.}~\bibnamefont {Laflamme}}, \bibinfo
  {author} {\bibfnamefont {Y.}~\bibnamefont {Nakamura}}, \bibinfo {author}
  {\bibfnamefont {C.}~\bibnamefont {Monroe}},\ and\ \bibinfo {author}
  {\bibfnamefont {J.~L.}\ \bibnamefont {O'Brien}},\ }\bibfield  {title}
  {\bibinfo {title} {Quantum computers},\ }\href
  {https://doi.org/10.1038/nature08812} {\bibfield  {journal} {\bibinfo
  {journal} {Nature (London)}\ }\textbf {\bibinfo {volume} {464}},\ \bibinfo
  {pages} {45} (\bibinfo {year} {2010})}\BibitemShut {NoStop}%
\bibitem [{\citenamefont {Monz}\ \emph {et~al.}(2016)\citenamefont {Monz},
  \citenamefont {Nigg}, \citenamefont {Martinez}, \citenamefont {Brandl},
  \citenamefont {Schindler}, \citenamefont {Rines}, \citenamefont {Wang},
  \citenamefont {Chuang},\ and\ \citenamefont {Blatt}}]{Monz2016}%
  \BibitemOpen
  \bibfield  {author} {\bibinfo {author} {\bibfnamefont {T.}~\bibnamefont
  {Monz}}, \bibinfo {author} {\bibfnamefont {D.}~\bibnamefont {Nigg}}, \bibinfo
  {author} {\bibfnamefont {E.~A.}\ \bibnamefont {Martinez}}, \bibinfo {author}
  {\bibfnamefont {M.~F.}\ \bibnamefont {Brandl}}, \bibinfo {author}
  {\bibfnamefont {P.}~\bibnamefont {Schindler}}, \bibinfo {author}
  {\bibfnamefont {R.}~\bibnamefont {Rines}}, \bibinfo {author} {\bibfnamefont
  {S.~X.}\ \bibnamefont {Wang}}, \bibinfo {author} {\bibfnamefont {I.~L.}\
  \bibnamefont {Chuang}},\ and\ \bibinfo {author} {\bibfnamefont
  {R.}~\bibnamefont {Blatt}},\ }\bibfield  {title} {\bibinfo {title}
  {Realization of a scalable shor algorithm},\ }\href
  {https://doi.org/10.1126/science.aad9480} {\bibfield  {journal} {\bibinfo
  {journal} {Science}\ }\textbf {\bibinfo {volume} {351}},\ \bibinfo {pages}
  {1068} (\bibinfo {year} {2016})}\BibitemShut {NoStop}%
\bibitem [{\citenamefont {Zhong}\ \emph {et~al.}(2020)\citenamefont {Zhong},
  \citenamefont {Wang}, \citenamefont {Deng}, \citenamefont {Chen},
  \citenamefont {Peng}, \citenamefont {Luo}, \citenamefont {Qin}, \citenamefont
  {Wu}, \citenamefont {Ding}, \citenamefont {Hu}, \citenamefont {Hu},
  \citenamefont {Yang}, \citenamefont {Zhang}, \citenamefont {Li},
  \citenamefont {Li}, \citenamefont {Jiang}, \citenamefont {Gan}, \citenamefont
  {Yang}, \citenamefont {You}, \citenamefont {Wang}, \citenamefont {Li},
  \citenamefont {Liu}, \citenamefont {Lu},\ and\ \citenamefont
  {Pan}}]{Zhong2020}%
  \BibitemOpen
  \bibfield  {author} {\bibinfo {author} {\bibfnamefont {H.-S.}\ \bibnamefont
  {Zhong}}, \bibinfo {author} {\bibfnamefont {H.}~\bibnamefont {Wang}},
  \bibinfo {author} {\bibfnamefont {Y.-H.}\ \bibnamefont {Deng}}, \bibinfo
  {author} {\bibfnamefont {M.-C.}\ \bibnamefont {Chen}}, \bibinfo {author}
  {\bibfnamefont {L.-C.}\ \bibnamefont {Peng}}, \bibinfo {author}
  {\bibfnamefont {Y.-H.}\ \bibnamefont {Luo}}, \bibinfo {author} {\bibfnamefont
  {J.}~\bibnamefont {Qin}}, \bibinfo {author} {\bibfnamefont {D.}~\bibnamefont
  {Wu}}, \bibinfo {author} {\bibfnamefont {X.}~\bibnamefont {Ding}}, \bibinfo
  {author} {\bibfnamefont {Y.}~\bibnamefont {Hu}}, \bibinfo {author}
  {\bibfnamefont {P.}~\bibnamefont {Hu}}, \bibinfo {author} {\bibfnamefont
  {X.-Y.}\ \bibnamefont {Yang}}, \bibinfo {author} {\bibfnamefont {W.-J.}\
  \bibnamefont {Zhang}}, \bibinfo {author} {\bibfnamefont {H.}~\bibnamefont
  {Li}}, \bibinfo {author} {\bibfnamefont {Y.}~\bibnamefont {Li}}, \bibinfo
  {author} {\bibfnamefont {X.}~\bibnamefont {Jiang}}, \bibinfo {author}
  {\bibfnamefont {L.}~\bibnamefont {Gan}}, \bibinfo {author} {\bibfnamefont
  {G.}~\bibnamefont {Yang}}, \bibinfo {author} {\bibfnamefont {L.}~\bibnamefont
  {You}}, \bibinfo {author} {\bibfnamefont {Z.}~\bibnamefont {Wang}}, \bibinfo
  {author} {\bibfnamefont {L.}~\bibnamefont {Li}}, \bibinfo {author}
  {\bibfnamefont {N.-L.}\ \bibnamefont {Liu}}, \bibinfo {author} {\bibfnamefont
  {C.-Y.}\ \bibnamefont {Lu}},\ and\ \bibinfo {author} {\bibfnamefont {J.-W.}\
  \bibnamefont {Pan}},\ }\bibfield  {title} {\bibinfo {title} {Quantum
  computational advantage using photons},\ }\href
  {https://doi.org/10.1126/science.abe8770} {\bibfield  {journal} {\bibinfo
  {journal} {Science}\ }\textbf {\bibinfo {volume} {370}},\ \bibinfo {pages}
  {1460} (\bibinfo {year} {2020})}\BibitemShut {NoStop}%
\bibitem [{\citenamefont {Nielsen}\ and\ \citenamefont
  {Chuang}(2004)}]{chuang}%
  \BibitemOpen
  \bibfield  {author} {\bibinfo {author} {\bibfnamefont {M.~A.}\ \bibnamefont
  {Nielsen}}\ and\ \bibinfo {author} {\bibfnamefont {I.}~\bibnamefont
  {Chuang}},\ }\href@noop {} {\emph {\bibinfo {title} {Quantum computation and
  quantum information}}}\ (\bibinfo  {publisher} {Cambridge University Press,
  Cambridge},\ \bibinfo {year} {2004})\BibitemShut {NoStop}%
\bibitem [{\citenamefont {Smolin}\ and\ \citenamefont
  {DiVincenzo}(1996)}]{PhysRevA.53.2855}%
  \BibitemOpen
  \bibfield  {author} {\bibinfo {author} {\bibfnamefont {J.~A.}\ \bibnamefont
  {Smolin}}\ and\ \bibinfo {author} {\bibfnamefont {D.~P.}\ \bibnamefont
  {DiVincenzo}},\ }\bibfield  {title} {\bibinfo {title} {Five two-bit quantum
  gates are sufficient to implement the quantum fredkin gate},\ }\href
  {https://doi.org/10.1103/PhysRevA.53.2855} {\bibfield  {journal} {\bibinfo
  {journal} {Phys. Rev. A}\ }\textbf {\bibinfo {volume} {53}},\ \bibinfo
  {pages} {2855} (\bibinfo {year} {1996})}\BibitemShut {NoStop}%
\bibitem [{\citenamefont {Vandersypen}\ \emph {et~al.}(2001)\citenamefont
  {Vandersypen}, \citenamefont {Steffen}, \citenamefont {Breyta}, \citenamefont
  {Yannoni}, \citenamefont {Sherwood},\ and\ \citenamefont
  {Chuang}}]{Vandersypen2001}%
  \BibitemOpen
  \bibfield  {author} {\bibinfo {author} {\bibfnamefont {L.~M.~K.}\
  \bibnamefont {Vandersypen}}, \bibinfo {author} {\bibfnamefont
  {M.}~\bibnamefont {Steffen}}, \bibinfo {author} {\bibfnamefont
  {G.}~\bibnamefont {Breyta}}, \bibinfo {author} {\bibfnamefont {C.~S.}\
  \bibnamefont {Yannoni}}, \bibinfo {author} {\bibfnamefont {M.~H.}\
  \bibnamefont {Sherwood}},\ and\ \bibinfo {author} {\bibfnamefont {I.~L.}\
  \bibnamefont {Chuang}},\ }\bibfield  {title} {\bibinfo {title} {Experimental
  realization of shor{\textquotesingle}s quantum factoring algorithm using
  nuclear magnetic resonance},\ }\href {https://doi.org/10.1038/414883a}
  {\bibfield  {journal} {\bibinfo  {journal} {Nature (London)}\ }\textbf
  {\bibinfo {volume} {414}},\ \bibinfo {pages} {883} (\bibinfo {year}
  {2001})}\BibitemShut {NoStop}%
\bibitem [{\citenamefont {Mart{\'{\i}}n-L{\'{o}}pez}\ \emph
  {et~al.}(2012)\citenamefont {Mart{\'{\i}}n-L{\'{o}}pez}, \citenamefont
  {Laing}, \citenamefont {Lawson}, \citenamefont {Alvarez}, \citenamefont
  {Zhou},\ and\ \citenamefont {O{\textquotesingle}Brien}}]{MartnLpez2012}%
  \BibitemOpen
  \bibfield  {author} {\bibinfo {author} {\bibfnamefont {E.}~\bibnamefont
  {Mart{\'{\i}}n-L{\'{o}}pez}}, \bibinfo {author} {\bibfnamefont
  {A.}~\bibnamefont {Laing}}, \bibinfo {author} {\bibfnamefont
  {T.}~\bibnamefont {Lawson}}, \bibinfo {author} {\bibfnamefont
  {R.}~\bibnamefont {Alvarez}}, \bibinfo {author} {\bibfnamefont {X.-Q.}\
  \bibnamefont {Zhou}},\ and\ \bibinfo {author} {\bibfnamefont {J.~L.}\
  \bibnamefont {O{\textquotesingle}Brien}},\ }\bibfield  {title} {\bibinfo
  {title} {Experimental realization of shor{\textquotesingle}s quantum
  factoring algorithm using qubit recycling},\ }\href
  {https://doi.org/10.1038/nphoton.2012.259} {\bibfield  {journal} {\bibinfo
  {journal} {Nat. Photonics}\ }\textbf {\bibinfo {volume} {6}},\ \bibinfo
  {pages} {773} (\bibinfo {year} {2012})}\BibitemShut {NoStop}%
\bibitem [{\citenamefont {Lanyon}\ \emph {et~al.}(2007)\citenamefont {Lanyon},
  \citenamefont {Weinhold}, \citenamefont {Langford}, \citenamefont {Barbieri},
  \citenamefont {James}, \citenamefont {Gilchrist},\ and\ \citenamefont
  {White}}]{PhysRevLett.99.250505}%
  \BibitemOpen
  \bibfield  {author} {\bibinfo {author} {\bibfnamefont {B.~P.}\ \bibnamefont
  {Lanyon}}, \bibinfo {author} {\bibfnamefont {T.~J.}\ \bibnamefont
  {Weinhold}}, \bibinfo {author} {\bibfnamefont {N.~K.}\ \bibnamefont
  {Langford}}, \bibinfo {author} {\bibfnamefont {M.}~\bibnamefont {Barbieri}},
  \bibinfo {author} {\bibfnamefont {D.~F.~V.}\ \bibnamefont {James}}, \bibinfo
  {author} {\bibfnamefont {A.}~\bibnamefont {Gilchrist}},\ and\ \bibinfo
  {author} {\bibfnamefont {A.~G.}\ \bibnamefont {White}},\ }\bibfield  {title}
  {\bibinfo {title} {{Experimental Demonstration of a Compiled Version of
  Shor's Algorithm with Quantum Entanglement}},\ }\href
  {https://doi.org/10.1103/PhysRevLett.99.250505} {\bibfield  {journal}
  {\bibinfo  {journal} {Phys. Rev. Lett.}\ }\textbf {\bibinfo {volume} {99}},\
  \bibinfo {pages} {250505} (\bibinfo {year} {2007})}\BibitemShut {NoStop}%
\bibitem [{\citenamefont {Buhrman}\ \emph {et~al.}(2001)\citenamefont
  {Buhrman}, \citenamefont {Cleve}, \citenamefont {Watrous},\ and\
  \citenamefont {de~Wolf}}]{PhysRevLett.87.167902}%
  \BibitemOpen
  \bibfield  {author} {\bibinfo {author} {\bibfnamefont {H.}~\bibnamefont
  {Buhrman}}, \bibinfo {author} {\bibfnamefont {R.}~\bibnamefont {Cleve}},
  \bibinfo {author} {\bibfnamefont {J.}~\bibnamefont {Watrous}},\ and\ \bibinfo
  {author} {\bibfnamefont {R.}~\bibnamefont {de~Wolf}},\ }\bibfield  {title}
  {\bibinfo {title} {Quantum {F}ingerprinting},\ }\href
  {https://doi.org/10.1103/PhysRevLett.87.167902} {\bibfield  {journal}
  {\bibinfo  {journal} {Phys. Rev. Lett.}\ }\textbf {\bibinfo {volume} {87}},\
  \bibinfo {pages} {167902} (\bibinfo {year} {2001})}\BibitemShut {NoStop}%
\bibitem [{\citenamefont {Horn}\ \emph {et~al.}(2005)\citenamefont {Horn},
  \citenamefont {Babichev}, \citenamefont {Marzlin}, \citenamefont {Lvovsky},\
  and\ \citenamefont {Sanders}}]{PhysRevLett.95.150502}%
  \BibitemOpen
  \bibfield  {author} {\bibinfo {author} {\bibfnamefont {R.~T.}\ \bibnamefont
  {Horn}}, \bibinfo {author} {\bibfnamefont {S.~A.}\ \bibnamefont {Babichev}},
  \bibinfo {author} {\bibfnamefont {K.-P.}\ \bibnamefont {Marzlin}}, \bibinfo
  {author} {\bibfnamefont {A.~I.}\ \bibnamefont {Lvovsky}},\ and\ \bibinfo
  {author} {\bibfnamefont {B.~C.}\ \bibnamefont {Sanders}},\ }\bibfield
  {title} {\bibinfo {title} {{Single-Qubit Optical Quantum Fingerprinting}},\
  }\href {https://doi.org/10.1103/PhysRevLett.95.150502} {\bibfield  {journal}
  {\bibinfo  {journal} {Phys. Rev. Lett.}\ }\textbf {\bibinfo {volume} {95}},\
  \bibinfo {pages} {150502} (\bibinfo {year} {2005})}\BibitemShut {NoStop}%
\bibitem [{\citenamefont {Chuang}\ and\ \citenamefont
  {Yamamoto}(1996)}]{PhysRevLett.76.4281}%
  \BibitemOpen
  \bibfield  {author} {\bibinfo {author} {\bibfnamefont {I.~L.}\ \bibnamefont
  {Chuang}}\ and\ \bibinfo {author} {\bibfnamefont {Y.}~\bibnamefont
  {Yamamoto}},\ }\bibfield  {title} {\bibinfo {title} {Quantum bit
  regeneration},\ }\href {https://doi.org/10.1103/PhysRevLett.76.4281}
  {\bibfield  {journal} {\bibinfo  {journal} {Phys. Rev. Lett.}\ }\textbf
  {\bibinfo {volume} {76}},\ \bibinfo {pages} {4281} (\bibinfo {year}
  {1996})}\BibitemShut {NoStop}%
\bibitem [{\citenamefont {Barenco}\ \emph {et~al.}(1997)\citenamefont
  {Barenco}, \citenamefont {Berthiaume}, \citenamefont {Deutsch}, \citenamefont
  {Ekert}, \citenamefont {Jozsa},\ and\ \citenamefont
  {Macchiavello}}]{Barenco1997}%
  \BibitemOpen
  \bibfield  {author} {\bibinfo {author} {\bibfnamefont {A.}~\bibnamefont
  {Barenco}}, \bibinfo {author} {\bibfnamefont {A.}~\bibnamefont {Berthiaume}},
  \bibinfo {author} {\bibfnamefont {D.}~\bibnamefont {Deutsch}}, \bibinfo
  {author} {\bibfnamefont {A.}~\bibnamefont {Ekert}}, \bibinfo {author}
  {\bibfnamefont {R.}~\bibnamefont {Jozsa}},\ and\ \bibinfo {author}
  {\bibfnamefont {C.}~\bibnamefont {Macchiavello}},\ }\bibfield  {title}
  {\bibinfo {title} {{Stabilization of Quantum Computations by
  Symmetrization}},\ }\href {https://doi.org/10.1137/s0097539796302452}
  {\bibfield  {journal} {\bibinfo  {journal} {{SIAM} J. Comp.}\ }\textbf
  {\bibinfo {volume} {26}},\ \bibinfo {pages} {1541} (\bibinfo {year}
  {1997})}\BibitemShut {NoStop}%
\bibitem [{\citenamefont {Ekert}\ \emph {et~al.}(2002)\citenamefont {Ekert},
  \citenamefont {Alves}, \citenamefont {Oi}, \citenamefont {Horodecki},
  \citenamefont {Horodecki},\ and\ \citenamefont
  {Kwek}}]{PhysRevLett.88.217901}%
  \BibitemOpen
  \bibfield  {author} {\bibinfo {author} {\bibfnamefont {A.~K.}\ \bibnamefont
  {Ekert}}, \bibinfo {author} {\bibfnamefont {C.~M.}\ \bibnamefont {Alves}},
  \bibinfo {author} {\bibfnamefont {D.~K.~L.}\ \bibnamefont {Oi}}, \bibinfo
  {author} {\bibfnamefont {M.}~\bibnamefont {Horodecki}}, \bibinfo {author}
  {\bibfnamefont {P.}~\bibnamefont {Horodecki}},\ and\ \bibinfo {author}
  {\bibfnamefont {L.~C.}\ \bibnamefont {Kwek}},\ }\bibfield  {title} {\bibinfo
  {title} {{Direct Estimations of Linear and Nonlinear Functionals of a Quantum
  State}},\ }\href {https://doi.org/10.1103/PhysRevLett.88.217901} {\bibfield
  {journal} {\bibinfo  {journal} {Phys. Rev. Lett.}\ }\textbf {\bibinfo
  {volume} {88}},\ \bibinfo {pages} {217901} (\bibinfo {year}
  {2002})}\BibitemShut {NoStop}%
\bibitem [{\citenamefont {Fiur\'a\ifmmode~\check{s}\else \v{s}\fi{}ek}\ \emph
  {et~al.}(2002)\citenamefont {Fiur\'a\ifmmode~\check{s}\else \v{s}\fi{}ek},
  \citenamefont {Du\ifmmode~\check{s}\else \v{s}\fi{}ek},\ and\ \citenamefont
  {Filip}}]{PhysRevLett.89.190401}%
  \BibitemOpen
  \bibfield  {author} {\bibinfo {author} {\bibfnamefont {J.}~\bibnamefont
  {Fiur\'a\ifmmode~\check{s}\else \v{s}\fi{}ek}}, \bibinfo {author}
  {\bibfnamefont {M.}~\bibnamefont {Du\ifmmode~\check{s}\else \v{s}\fi{}ek}},\
  and\ \bibinfo {author} {\bibfnamefont {R.}~\bibnamefont {Filip}},\ }\bibfield
   {title} {\bibinfo {title} {{Universal Measurement Apparatus Controlled by
  Quantum Software}},\ }\href {https://doi.org/10.1103/PhysRevLett.89.190401}
  {\bibfield  {journal} {\bibinfo  {journal} {Phys. Rev. Lett.}\ }\textbf
  {\bibinfo {volume} {89}},\ \bibinfo {pages} {190401} (\bibinfo {year}
  {2002})}\BibitemShut {NoStop}%
\bibitem [{\citenamefont {Patel}\ \emph {et~al.}(2016)\citenamefont {Patel},
  \citenamefont {Ho}, \citenamefont {Ferreyrol}, \citenamefont {Ralph},\ and\
  \citenamefont {Pryde}}]{Patel2016}%
  \BibitemOpen
  \bibfield  {author} {\bibinfo {author} {\bibfnamefont {R.~B.}\ \bibnamefont
  {Patel}}, \bibinfo {author} {\bibfnamefont {J.}~\bibnamefont {Ho}}, \bibinfo
  {author} {\bibfnamefont {F.}~\bibnamefont {Ferreyrol}}, \bibinfo {author}
  {\bibfnamefont {T.~C.}\ \bibnamefont {Ralph}},\ and\ \bibinfo {author}
  {\bibfnamefont {G.~J.}\ \bibnamefont {Pryde}},\ }\bibfield  {title} {\bibinfo
  {title} {A quantum {Fredkin} gate},\ }\href
  {https://doi.org/10.1126/sciadv.1501531} {\bibfield  {journal} {\bibinfo
  {journal} {Sci. Adv.}\ }\textbf {\bibinfo {volume} {2}} (\bibinfo {year}
  {2016})}\BibitemShut {NoStop}%
\bibitem [{\citenamefont {Li}\ \emph {et~al.}(2022)\citenamefont {Li},
  \citenamefont {Wan}, \citenamefont {Zhang}, \citenamefont {Zhu},
  \citenamefont {Shi}, \citenamefont {Chin}, \citenamefont {Zhou},
  \citenamefont {Kwek},\ and\ \citenamefont {Liu}}]{Li2022}%
  \BibitemOpen
  \bibfield  {author} {\bibinfo {author} {\bibfnamefont {Y.}~\bibnamefont
  {Li}}, \bibinfo {author} {\bibfnamefont {L.}~\bibnamefont {Wan}}, \bibinfo
  {author} {\bibfnamefont {H.}~\bibnamefont {Zhang}}, \bibinfo {author}
  {\bibfnamefont {H.}~\bibnamefont {Zhu}}, \bibinfo {author} {\bibfnamefont
  {Y.}~\bibnamefont {Shi}}, \bibinfo {author} {\bibfnamefont {L.~K.}\
  \bibnamefont {Chin}}, \bibinfo {author} {\bibfnamefont {X.}~\bibnamefont
  {Zhou}}, \bibinfo {author} {\bibfnamefont {L.~C.}\ \bibnamefont {Kwek}},\
  and\ \bibinfo {author} {\bibfnamefont {A.~Q.}\ \bibnamefont {Liu}},\
  }\bibfield  {title} {\bibinfo {title} {Quantum fredkin and toffoli gates on a
  versatile programmable silicon photonic chip},\ }\href
  {https://doi.org/10.1038/s41534-022-00627-y} {\bibfield  {journal} {\bibinfo
  {journal} {NPJ Quantum Inf.}\ }\textbf {\bibinfo {volume} {8}} (\bibinfo
  {year} {2022})}\BibitemShut {NoStop}%
\bibitem [{\citenamefont {Gong}\ \emph {et~al.}(2021)\citenamefont {Gong},
  \citenamefont {Wang}, \citenamefont {Zha}, \citenamefont {Chen},
  \citenamefont {Huang}, \citenamefont {Wu}, \citenamefont {Zhu}, \citenamefont
  {Zhao}, \citenamefont {Li}, \citenamefont {Guo}, \citenamefont {Qian},
  \citenamefont {Ye}, \citenamefont {Chen}, \citenamefont {Ying}, \citenamefont
  {Yu}, \citenamefont {Fan}, \citenamefont {Wu}, \citenamefont {Su},
  \citenamefont {Deng}, \citenamefont {Rong}, \citenamefont {Zhang},
  \citenamefont {Cao}, \citenamefont {Lin}, \citenamefont {Xu}, \citenamefont
  {Sun}, \citenamefont {Guo}, \citenamefont {Li}, \citenamefont {Liang},
  \citenamefont {Bastidas}, \citenamefont {Nemoto}, \citenamefont {Munro},
  \citenamefont {Huo}, \citenamefont {Lu}, \citenamefont {Peng}, \citenamefont
  {Zhu},\ and\ \citenamefont {Pan}}]{Gong2021}%
  \BibitemOpen
  \bibfield  {author} {\bibinfo {author} {\bibfnamefont {M.}~\bibnamefont
  {Gong}}, \bibinfo {author} {\bibfnamefont {S.}~\bibnamefont {Wang}}, \bibinfo
  {author} {\bibfnamefont {C.}~\bibnamefont {Zha}}, \bibinfo {author}
  {\bibfnamefont {M.-C.}\ \bibnamefont {Chen}}, \bibinfo {author}
  {\bibfnamefont {H.-L.}\ \bibnamefont {Huang}}, \bibinfo {author}
  {\bibfnamefont {Y.}~\bibnamefont {Wu}}, \bibinfo {author} {\bibfnamefont
  {Q.}~\bibnamefont {Zhu}}, \bibinfo {author} {\bibfnamefont {Y.}~\bibnamefont
  {Zhao}}, \bibinfo {author} {\bibfnamefont {S.}~\bibnamefont {Li}}, \bibinfo
  {author} {\bibfnamefont {S.}~\bibnamefont {Guo}}, \bibinfo {author}
  {\bibfnamefont {H.}~\bibnamefont {Qian}}, \bibinfo {author} {\bibfnamefont
  {Y.}~\bibnamefont {Ye}}, \bibinfo {author} {\bibfnamefont {F.}~\bibnamefont
  {Chen}}, \bibinfo {author} {\bibfnamefont {C.}~\bibnamefont {Ying}}, \bibinfo
  {author} {\bibfnamefont {J.}~\bibnamefont {Yu}}, \bibinfo {author}
  {\bibfnamefont {D.}~\bibnamefont {Fan}}, \bibinfo {author} {\bibfnamefont
  {D.}~\bibnamefont {Wu}}, \bibinfo {author} {\bibfnamefont {H.}~\bibnamefont
  {Su}}, \bibinfo {author} {\bibfnamefont {H.}~\bibnamefont {Deng}}, \bibinfo
  {author} {\bibfnamefont {H.}~\bibnamefont {Rong}}, \bibinfo {author}
  {\bibfnamefont {K.}~\bibnamefont {Zhang}}, \bibinfo {author} {\bibfnamefont
  {S.}~\bibnamefont {Cao}}, \bibinfo {author} {\bibfnamefont {J.}~\bibnamefont
  {Lin}}, \bibinfo {author} {\bibfnamefont {Y.}~\bibnamefont {Xu}}, \bibinfo
  {author} {\bibfnamefont {L.}~\bibnamefont {Sun}}, \bibinfo {author}
  {\bibfnamefont {C.}~\bibnamefont {Guo}}, \bibinfo {author} {\bibfnamefont
  {N.}~\bibnamefont {Li}}, \bibinfo {author} {\bibfnamefont {F.}~\bibnamefont
  {Liang}}, \bibinfo {author} {\bibfnamefont {V.~M.}\ \bibnamefont {Bastidas}},
  \bibinfo {author} {\bibfnamefont {K.}~\bibnamefont {Nemoto}}, \bibinfo
  {author} {\bibfnamefont {W.~J.}\ \bibnamefont {Munro}}, \bibinfo {author}
  {\bibfnamefont {Y.-H.}\ \bibnamefont {Huo}}, \bibinfo {author} {\bibfnamefont
  {C.-Y.}\ \bibnamefont {Lu}}, \bibinfo {author} {\bibfnamefont {C.-Z.}\
  \bibnamefont {Peng}}, \bibinfo {author} {\bibfnamefont {X.}~\bibnamefont
  {Zhu}},\ and\ \bibinfo {author} {\bibfnamefont {J.-W.}\ \bibnamefont {Pan}},\
  }\bibfield  {title} {\bibinfo {title} {Quantum walks on a programmable
  two-dimensional 62-qubit superconducting processor},\ }\href
  {https://doi.org/10.1126/science.abg7812} {\bibfield  {journal} {\bibinfo
  {journal} {Science}\ }\textbf {\bibinfo {volume} {372}},\ \bibinfo {pages}
  {948} (\bibinfo {year} {2021})}\BibitemShut {NoStop}%
\bibitem [{\citenamefont {Carolan}\ \emph {et~al.}(2015)\citenamefont
  {Carolan}, \citenamefont {Harrold}, \citenamefont {Sparrow}, \citenamefont
  {Mart{\'{\i}}n-L{\'{o}}pez}, \citenamefont {Russell}, \citenamefont
  {Silverstone}, \citenamefont {Shadbolt}, \citenamefont {Matsuda},
  \citenamefont {Oguma}, \citenamefont {Itoh}, \citenamefont {Marshall},
  \citenamefont {Thompson}, \citenamefont {Matthews}, \citenamefont
  {Hashimoto}, \citenamefont {O'Brien},\ and\ \citenamefont
  {Laing}}]{Carolan2015}%
  \BibitemOpen
  \bibfield  {author} {\bibinfo {author} {\bibfnamefont {J.}~\bibnamefont
  {Carolan}}, \bibinfo {author} {\bibfnamefont {C.}~\bibnamefont {Harrold}},
  \bibinfo {author} {\bibfnamefont {C.}~\bibnamefont {Sparrow}}, \bibinfo
  {author} {\bibfnamefont {E.}~\bibnamefont {Mart{\'{\i}}n-L{\'{o}}pez}},
  \bibinfo {author} {\bibfnamefont {N.~J.}\ \bibnamefont {Russell}}, \bibinfo
  {author} {\bibfnamefont {J.~W.}\ \bibnamefont {Silverstone}}, \bibinfo
  {author} {\bibfnamefont {P.~J.}\ \bibnamefont {Shadbolt}}, \bibinfo {author}
  {\bibfnamefont {N.}~\bibnamefont {Matsuda}}, \bibinfo {author} {\bibfnamefont
  {M.}~\bibnamefont {Oguma}}, \bibinfo {author} {\bibfnamefont
  {M.}~\bibnamefont {Itoh}}, \bibinfo {author} {\bibfnamefont {G.~D.}\
  \bibnamefont {Marshall}}, \bibinfo {author} {\bibfnamefont {M.~G.}\
  \bibnamefont {Thompson}}, \bibinfo {author} {\bibfnamefont {J.~C.~F.}\
  \bibnamefont {Matthews}}, \bibinfo {author} {\bibfnamefont {T.}~\bibnamefont
  {Hashimoto}}, \bibinfo {author} {\bibfnamefont {J.~L.}\ \bibnamefont
  {O'Brien}},\ and\ \bibinfo {author} {\bibfnamefont {A.}~\bibnamefont
  {Laing}},\ }\bibfield  {title} {\bibinfo {title} {Universal linear optics},\
  }\href {https://doi.org/10.1126/science.aab3642} {\bibfield  {journal}
  {\bibinfo  {journal} {Science}\ }\textbf {\bibinfo {volume} {349}},\ \bibinfo
  {pages} {711} (\bibinfo {year} {2015})}\BibitemShut {NoStop}%
\bibitem [{\citenamefont {Ebadi}\ \emph {et~al.}(2021)\citenamefont {Ebadi},
  \citenamefont {Wang}, \citenamefont {Levine}, \citenamefont {Keesling},
  \citenamefont {Semeghini}, \citenamefont {Omran}, \citenamefont {Bluvstein},
  \citenamefont {Samajdar}, \citenamefont {Pichler}, \citenamefont {Ho},
  \citenamefont {Choi}, \citenamefont {Sachdev}, \citenamefont {Greiner},
  \citenamefont {Vuleti{\'{c}}},\ and\ \citenamefont {Lukin}}]{Ebadi2021}%
  \BibitemOpen
  \bibfield  {author} {\bibinfo {author} {\bibfnamefont {S.}~\bibnamefont
  {Ebadi}}, \bibinfo {author} {\bibfnamefont {T.~T.}\ \bibnamefont {Wang}},
  \bibinfo {author} {\bibfnamefont {H.}~\bibnamefont {Levine}}, \bibinfo
  {author} {\bibfnamefont {A.}~\bibnamefont {Keesling}}, \bibinfo {author}
  {\bibfnamefont {G.}~\bibnamefont {Semeghini}}, \bibinfo {author}
  {\bibfnamefont {A.}~\bibnamefont {Omran}}, \bibinfo {author} {\bibfnamefont
  {D.}~\bibnamefont {Bluvstein}}, \bibinfo {author} {\bibfnamefont
  {R.}~\bibnamefont {Samajdar}}, \bibinfo {author} {\bibfnamefont
  {H.}~\bibnamefont {Pichler}}, \bibinfo {author} {\bibfnamefont {W.~W.}\
  \bibnamefont {Ho}}, \bibinfo {author} {\bibfnamefont {S.}~\bibnamefont
  {Choi}}, \bibinfo {author} {\bibfnamefont {S.}~\bibnamefont {Sachdev}},
  \bibinfo {author} {\bibfnamefont {M.}~\bibnamefont {Greiner}}, \bibinfo
  {author} {\bibfnamefont {V.}~\bibnamefont {Vuleti{\'{c}}}},\ and\ \bibinfo
  {author} {\bibfnamefont {M.~D.}\ \bibnamefont {Lukin}},\ }\bibfield  {title}
  {\bibinfo {title} {Quantum phases of matter on a 256-atom programmable
  quantum simulator},\ }\href {https://doi.org/10.1038/s41586-021-03582-4}
  {\bibfield  {journal} {\bibinfo  {journal} {Nature (London)}\ }\textbf
  {\bibinfo {volume} {595}},\ \bibinfo {pages} {227} (\bibinfo {year}
  {2021})}\BibitemShut {NoStop}%
\bibitem [{\citenamefont {Reiserer}\ and\ \citenamefont
  {Rempe}(2015)}]{rempe2015}%
  \BibitemOpen
  \bibfield  {author} {\bibinfo {author} {\bibfnamefont {A.}~\bibnamefont
  {Reiserer}}\ and\ \bibinfo {author} {\bibfnamefont {G.}~\bibnamefont
  {Rempe}},\ }\bibfield  {title} {\bibinfo {title} {Cavity-based quantum
  networks with single atoms and optical photons},\ }\href
  {https://doi.org/10.1103/RevModPhys.87.1379} {\bibfield  {journal} {\bibinfo
  {journal} {Rev. Mod. Phys.}\ }\textbf {\bibinfo {volume} {87}},\ \bibinfo
  {pages} {1379} (\bibinfo {year} {2015})}\BibitemShut {NoStop}%
\bibitem [{\citenamefont {Pellizzari}\ \emph {et~al.}(1995)\citenamefont
  {Pellizzari}, \citenamefont {Gardiner}, \citenamefont {Cirac},\ and\
  \citenamefont {Zoller}}]{zoller95}%
  \BibitemOpen
  \bibfield  {author} {\bibinfo {author} {\bibfnamefont {T.}~\bibnamefont
  {Pellizzari}}, \bibinfo {author} {\bibfnamefont {S.~A.}\ \bibnamefont
  {Gardiner}}, \bibinfo {author} {\bibfnamefont {J.~I.}\ \bibnamefont
  {Cirac}},\ and\ \bibinfo {author} {\bibfnamefont {P.}~\bibnamefont
  {Zoller}},\ }\bibfield  {title} {\bibinfo {title} {{Decoherence, Continuous
  Observation, and Quantum Computing: A Cavity QED Model}},\ }\href
  {https://doi.org/10.1103/PhysRevLett.75.3788} {\bibfield  {journal} {\bibinfo
   {journal} {Phys. Rev. Lett.}\ }\textbf {\bibinfo {volume} {75}},\ \bibinfo
  {pages} {3788} (\bibinfo {year} {1995})}\BibitemShut {NoStop}%
\bibitem [{\citenamefont {Cirac}\ \emph {et~al.}(1997)\citenamefont {Cirac},
  \citenamefont {Zoller}, \citenamefont {Kimble},\ and\ \citenamefont
  {Mabuchi}}]{cirac97}%
  \BibitemOpen
  \bibfield  {author} {\bibinfo {author} {\bibfnamefont {J.~I.}\ \bibnamefont
  {Cirac}}, \bibinfo {author} {\bibfnamefont {P.}~\bibnamefont {Zoller}},
  \bibinfo {author} {\bibfnamefont {H.~J.}\ \bibnamefont {Kimble}},\ and\
  \bibinfo {author} {\bibfnamefont {H.}~\bibnamefont {Mabuchi}},\ }\bibfield
  {title} {\bibinfo {title} {{Quantum State Transfer and Entanglement
  Distribution among Distant Nodes in a Quantum Network}},\ }\href
  {https://doi.org/10.1103/PhysRevLett.78.3221} {\bibfield  {journal} {\bibinfo
   {journal} {Phys. Rev. Lett.}\ }\textbf {\bibinfo {volume} {78}},\ \bibinfo
  {pages} {3221} (\bibinfo {year} {1997})}\BibitemShut {NoStop}%
\bibitem [{\citenamefont {Duan}\ and\ \citenamefont {Kimble}(2004)}]{duan04}%
  \BibitemOpen
  \bibfield  {author} {\bibinfo {author} {\bibfnamefont {L.-M.}\ \bibnamefont
  {Duan}}\ and\ \bibinfo {author} {\bibfnamefont {H.~J.}\ \bibnamefont
  {Kimble}},\ }\bibfield  {title} {\bibinfo {title} {{Scalable Photonic Quantum
  Computation through Cavity-Assisted Interactions}},\ }\href
  {https://doi.org/10.1103/PhysRevLett.92.127902} {\bibfield  {journal}
  {\bibinfo  {journal} {Phys. Rev. Lett.}\ }\textbf {\bibinfo {volume} {92}},\
  \bibinfo {pages} {127902} (\bibinfo {year} {2004})}\BibitemShut {NoStop}%
\bibitem [{\citenamefont {Kimble}(2008)}]{kimbleqi}%
  \BibitemOpen
  \bibfield  {author} {\bibinfo {author} {\bibfnamefont {H.~J.}\ \bibnamefont
  {Kimble}},\ }\bibfield  {title} {\bibinfo {title} {The quantum internet},\
  }\href {https://doi.org/10.1038/nature07127} {\bibfield  {journal} {\bibinfo
  {journal} {Nature (London)}\ }\textbf {\bibinfo {volume} {453}},\ \bibinfo
  {pages} {1023} (\bibinfo {year} {2008})}\BibitemShut {NoStop}%
\bibitem [{\citenamefont {Monroe}(2002)}]{monroerev}%
  \BibitemOpen
  \bibfield  {author} {\bibinfo {author} {\bibfnamefont {C.}~\bibnamefont
  {Monroe}},\ }\bibfield  {title} {\bibinfo {title} {Quantum information
  processing with atoms and photons},\ }\href {https://doi.org/10.1038/416238a}
  {\bibfield  {journal} {\bibinfo  {journal} {Nature (London)}\ }\textbf
  {\bibinfo {volume} {416}},\ \bibinfo {pages} {9} (\bibinfo {year}
  {2002})}\BibitemShut {NoStop}%
\bibitem [{\citenamefont {Ritter}\ \emph {et~al.}(2012)\citenamefont {Ritter},
  \citenamefont {Nölleke}, \citenamefont {Hahn}, \citenamefont {Reiserer},
  \citenamefont {Neuzner}, \citenamefont {Uphoff}, \citenamefont {Mücke},
  \citenamefont {Figueroa}, \citenamefont {Bochmann},\ and\ \citenamefont
  {Rempe}}]{Ritter_2012}%
  \BibitemOpen
  \bibfield  {author} {\bibinfo {author} {\bibfnamefont {S.}~\bibnamefont
  {Ritter}}, \bibinfo {author} {\bibfnamefont {C.}~\bibnamefont {Nölleke}},
  \bibinfo {author} {\bibfnamefont {C.}~\bibnamefont {Hahn}}, \bibinfo {author}
  {\bibfnamefont {A.}~\bibnamefont {Reiserer}}, \bibinfo {author}
  {\bibfnamefont {A.}~\bibnamefont {Neuzner}}, \bibinfo {author} {\bibfnamefont
  {M.}~\bibnamefont {Uphoff}}, \bibinfo {author} {\bibfnamefont
  {M.}~\bibnamefont {Mücke}}, \bibinfo {author} {\bibfnamefont
  {E.}~\bibnamefont {Figueroa}}, \bibinfo {author} {\bibfnamefont
  {J.}~\bibnamefont {Bochmann}},\ and\ \bibinfo {author} {\bibfnamefont
  {G.}~\bibnamefont {Rempe}},\ }\bibfield  {title} {\bibinfo {title} {An
  elementary quantum network of single atoms in optical cavities},\ }\href
  {https://doi.org/10.1038/nature11023} {\bibfield  {journal} {\bibinfo
  {journal} {Nature (London)}\ }\textbf {\bibinfo {volume} {484}},\ \bibinfo
  {pages} {195} (\bibinfo {year} {2012})}\BibitemShut {NoStop}%
\bibitem [{\citenamefont {Brekenfeld}\ \emph {et~al.}(2020)\citenamefont
  {Brekenfeld}, \citenamefont {Niemietz}, \citenamefont {Christesen},\ and\
  \citenamefont {Rempe}}]{brekenfeld}%
  \BibitemOpen
  \bibfield  {author} {\bibinfo {author} {\bibfnamefont {M.}~\bibnamefont
  {Brekenfeld}}, \bibinfo {author} {\bibfnamefont {D.}~\bibnamefont
  {Niemietz}}, \bibinfo {author} {\bibfnamefont {J.~D.}\ \bibnamefont
  {Christesen}},\ and\ \bibinfo {author} {\bibfnamefont {G.}~\bibnamefont
  {Rempe}},\ }\bibfield  {title} {\bibinfo {title} {A quantum network node with
  crossed optical fibre cavities},\ }\href
  {https://doi.org/10.1038/s41567-020-0855-3} {\bibfield  {journal} {\bibinfo
  {journal} {Nat. Phys.}\ }\textbf {\bibinfo {volume} {16}},\ \bibinfo {pages}
  {647} (\bibinfo {year} {2020})}\BibitemShut {NoStop}%
\bibitem [{\citenamefont {Scarani}\ \emph {et~al.}(2009)\citenamefont
  {Scarani}, \citenamefont {Bechmann-Pasquinucci}, \citenamefont {Cerf},
  \citenamefont {Du\ifmmode~\check{s}\else \v{s}\fi{}ek}, \citenamefont
  {L\"utkenhaus},\ and\ \citenamefont {Peev}}]{scarani09}%
  \BibitemOpen
  \bibfield  {author} {\bibinfo {author} {\bibfnamefont {V.}~\bibnamefont
  {Scarani}}, \bibinfo {author} {\bibfnamefont {H.}~\bibnamefont
  {Bechmann-Pasquinucci}}, \bibinfo {author} {\bibfnamefont {N.~J.}\
  \bibnamefont {Cerf}}, \bibinfo {author} {\bibfnamefont {M.}~\bibnamefont
  {Du\ifmmode~\check{s}\else \v{s}\fi{}ek}}, \bibinfo {author} {\bibfnamefont
  {N.}~\bibnamefont {L\"utkenhaus}},\ and\ \bibinfo {author} {\bibfnamefont
  {M.}~\bibnamefont {Peev}},\ }\bibfield  {title} {\bibinfo {title} {The
  security of practical quantum key distribution},\ }\href
  {https://doi.org/10.1103/RevModPhys.81.1301} {\bibfield  {journal} {\bibinfo
  {journal} {Rev. Mod. Phys.}\ }\textbf {\bibinfo {volume} {81}},\ \bibinfo
  {pages} {1301} (\bibinfo {year} {2009})}\BibitemShut {NoStop}%
\bibitem [{\citenamefont {Langenfeld}\ \emph
  {et~al.}(2021{\natexlab{a}})\citenamefont {Langenfeld}, \citenamefont
  {Thomas}, \citenamefont {Morin},\ and\ \citenamefont {Rempe}}]{qrepeater}%
  \BibitemOpen
  \bibfield  {author} {\bibinfo {author} {\bibfnamefont {S.}~\bibnamefont
  {Langenfeld}}, \bibinfo {author} {\bibfnamefont {P.}~\bibnamefont {Thomas}},
  \bibinfo {author} {\bibfnamefont {O.}~\bibnamefont {Morin}},\ and\ \bibinfo
  {author} {\bibfnamefont {G.}~\bibnamefont {Rempe}},\ }\bibfield  {title}
  {\bibinfo {title} {{Quantum Repeater Node Demonstrating Unconditionally
  Secure Key Distribution}},\ }\href
  {https://doi.org/10.1103/PhysRevLett.126.230506} {\bibfield  {journal}
  {\bibinfo  {journal} {Phys. Rev. Lett.}\ }\textbf {\bibinfo {volume} {126}},\
  \bibinfo {pages} {230506} (\bibinfo {year} {2021}{\natexlab{a}})}\BibitemShut
  {NoStop}%
\bibitem [{\citenamefont {Choi}\ \emph {et~al.}(2008)\citenamefont {Choi},
  \citenamefont {Deng}, \citenamefont {Laurat},\ and\ \citenamefont
  {Kimble}}]{kimble08qm}%
  \BibitemOpen
  \bibfield  {author} {\bibinfo {author} {\bibfnamefont {K.~S.}\ \bibnamefont
  {Choi}}, \bibinfo {author} {\bibfnamefont {H.}~\bibnamefont {Deng}}, \bibinfo
  {author} {\bibfnamefont {J.}~\bibnamefont {Laurat}},\ and\ \bibinfo {author}
  {\bibfnamefont {H.~J.}\ \bibnamefont {Kimble}},\ }\bibfield  {title}
  {\bibinfo {title} {Mapping photonic entanglement into and out of a quantum
  memory},\ }\href {https://doi.org/10.1038/nature06670} {\bibfield  {journal}
  {\bibinfo  {journal} {Nature (London)}\ }\textbf {\bibinfo {volume} {453}},\
  \bibinfo {pages} {67} (\bibinfo {year} {2008})}\BibitemShut {NoStop}%
\bibitem [{\citenamefont {Langenfeld}\ \emph {et~al.}(2020)\citenamefont
  {Langenfeld}, \citenamefont {Morin}, \citenamefont {K\"{o}rber},\ and\
  \citenamefont {Rempe}}]{Langenfeld2020}%
  \BibitemOpen
  \bibfield  {author} {\bibinfo {author} {\bibfnamefont {S.}~\bibnamefont
  {Langenfeld}}, \bibinfo {author} {\bibfnamefont {O.}~\bibnamefont {Morin}},
  \bibinfo {author} {\bibfnamefont {M.}~\bibnamefont {K\"{o}rber}},\ and\
  \bibinfo {author} {\bibfnamefont {G.}~\bibnamefont {Rempe}},\ }\bibfield
  {title} {\bibinfo {title} {A network-ready random-access qubits memory},\
  }\href {https://doi.org/10.1038/s41534-020-00316-8} {\bibfield  {journal}
  {\bibinfo  {journal} {NPJ Quantum Inf.}\ }\textbf {\bibinfo {volume} {6}}
  (\bibinfo {year} {2020})}\BibitemShut {NoStop}%
\bibitem [{\citenamefont {Langenfeld}\ \emph
  {et~al.}(2021{\natexlab{b}})\citenamefont {Langenfeld}, \citenamefont
  {Welte}, \citenamefont {Hartung}, \citenamefont {Daiss}, \citenamefont
  {Thomas}, \citenamefont {Morin}, \citenamefont {Distante},\ and\
  \citenamefont {Rempe}}]{PhysRevLett.126.130502}%
  \BibitemOpen
  \bibfield  {author} {\bibinfo {author} {\bibfnamefont {S.}~\bibnamefont
  {Langenfeld}}, \bibinfo {author} {\bibfnamefont {S.}~\bibnamefont {Welte}},
  \bibinfo {author} {\bibfnamefont {L.}~\bibnamefont {Hartung}}, \bibinfo
  {author} {\bibfnamefont {S.}~\bibnamefont {Daiss}}, \bibinfo {author}
  {\bibfnamefont {P.}~\bibnamefont {Thomas}}, \bibinfo {author} {\bibfnamefont
  {O.}~\bibnamefont {Morin}}, \bibinfo {author} {\bibfnamefont
  {E.}~\bibnamefont {Distante}},\ and\ \bibinfo {author} {\bibfnamefont
  {G.}~\bibnamefont {Rempe}},\ }\bibfield  {title} {\bibinfo {title} {Quantum
  teleportation between remote qubit memories with only a single photon as a
  resource},\ }\href {https://doi.org/10.1103/PhysRevLett.126.130502}
  {\bibfield  {journal} {\bibinfo  {journal} {Phys. Rev. Lett.}\ }\textbf
  {\bibinfo {volume} {126}},\ \bibinfo {pages} {130502} (\bibinfo {year}
  {2021}{\natexlab{b}})}\BibitemShut {NoStop}%
\bibitem [{\citenamefont {Tiarks}\ \emph {et~al.}(2014)\citenamefont {Tiarks},
  \citenamefont {Baur}, \citenamefont {Schneider}, \citenamefont {D\"urr},\
  and\ \citenamefont {Rempe}}]{tiarks2014qt}%
  \BibitemOpen
  \bibfield  {author} {\bibinfo {author} {\bibfnamefont {D.}~\bibnamefont
  {Tiarks}}, \bibinfo {author} {\bibfnamefont {S.}~\bibnamefont {Baur}},
  \bibinfo {author} {\bibfnamefont {K.}~\bibnamefont {Schneider}}, \bibinfo
  {author} {\bibfnamefont {S.}~\bibnamefont {D\"urr}},\ and\ \bibinfo {author}
  {\bibfnamefont {G.}~\bibnamefont {Rempe}},\ }\bibfield  {title} {\bibinfo
  {title} {{Single-Photon Transistor Using a F\"orster Resonance}},\ }\href
  {https://doi.org/10.1103/PhysRevLett.113.053602} {\bibfield  {journal}
  {\bibinfo  {journal} {Phys. Rev. Lett.}\ }\textbf {\bibinfo {volume} {113}},\
  \bibinfo {pages} {053602} (\bibinfo {year} {2014})}\BibitemShut {NoStop}%
\bibitem [{\citenamefont {Turchette}\ \emph {et~al.}(1995)\citenamefont
  {Turchette}, \citenamefont {Hood}, \citenamefont {Lange}, \citenamefont
  {Mabuchi},\ and\ \citenamefont {Kimble}}]{turchette95}%
  \BibitemOpen
  \bibfield  {author} {\bibinfo {author} {\bibfnamefont {Q.~A.}\ \bibnamefont
  {Turchette}}, \bibinfo {author} {\bibfnamefont {C.~J.}\ \bibnamefont {Hood}},
  \bibinfo {author} {\bibfnamefont {W.}~\bibnamefont {Lange}}, \bibinfo
  {author} {\bibfnamefont {H.}~\bibnamefont {Mabuchi}},\ and\ \bibinfo {author}
  {\bibfnamefont {H.~J.}\ \bibnamefont {Kimble}},\ }\bibfield  {title}
  {\bibinfo {title} {{Measurement of Conditional Phase Shifts for Quantum
  Logic}},\ }\href {https://doi.org/10.1103/PhysRevLett.75.4710} {\bibfield
  {journal} {\bibinfo  {journal} {Phys. Rev. Lett.}\ }\textbf {\bibinfo
  {volume} {75}},\ \bibinfo {pages} {4710} (\bibinfo {year}
  {1995})}\BibitemShut {NoStop}%
\bibitem [{\citenamefont {Duan}\ \emph {et~al.}(2005)\citenamefont {Duan},
  \citenamefont {Wang},\ and\ \citenamefont {Kimble}}]{duan05}%
  \BibitemOpen
  \bibfield  {author} {\bibinfo {author} {\bibfnamefont {L.-M.}\ \bibnamefont
  {Duan}}, \bibinfo {author} {\bibfnamefont {B.}~\bibnamefont {Wang}},\ and\
  \bibinfo {author} {\bibfnamefont {H.~J.}\ \bibnamefont {Kimble}},\ }\bibfield
   {title} {\bibinfo {title} {Robust quantum gates on neutral atoms with
  cavity-assisted photon scattering},\ }\href
  {https://doi.org/10.1103/PhysRevA.72.032333} {\bibfield  {journal} {\bibinfo
  {journal} {Phys. Rev. A}\ }\textbf {\bibinfo {volume} {72}},\ \bibinfo
  {pages} {032333} (\bibinfo {year} {2005})}\BibitemShut {NoStop}%
\bibitem [{\citenamefont {Zhou}\ and\ \citenamefont {Li}(2011)}]{zhou2011}%
  \BibitemOpen
  \bibfield  {author} {\bibinfo {author} {\bibfnamefont {Y.~L.}\ \bibnamefont
  {Zhou}}\ and\ \bibinfo {author} {\bibfnamefont {C.~Z.}\ \bibnamefont {Li}},\
  }\bibfield  {title} {\bibinfo {title} {Robust quantum gates via a photon
  triggering electromagnetically induced transparency},\ }\href
  {https://doi.org/10.1103/PhysRevA.84.044304} {\bibfield  {journal} {\bibinfo
  {journal} {Phys. Rev. A}\ }\textbf {\bibinfo {volume} {84}},\ \bibinfo
  {pages} {044304} (\bibinfo {year} {2011})}\BibitemShut {NoStop}%
\bibitem [{\citenamefont {Borges}\ and\ \citenamefont
  {Villas-B\^oas}(2016)}]{cvb16}%
  \BibitemOpen
  \bibfield  {author} {\bibinfo {author} {\bibfnamefont {H.~S.}\ \bibnamefont
  {Borges}}\ and\ \bibinfo {author} {\bibfnamefont {C.~J.}\ \bibnamefont
  {Villas-B\^oas}},\ }\bibfield  {title} {\bibinfo {title} {Quantum phase gate
  based on electromagnetically induced transparency in optical cavities},\
  }\href {https://doi.org/10.1103/PhysRevA.94.052337} {\bibfield  {journal}
  {\bibinfo  {journal} {Phys. Rev. A}\ }\textbf {\bibinfo {volume} {94}},\
  \bibinfo {pages} {052337} (\bibinfo {year} {2016})}\BibitemShut {NoStop}%
\bibitem [{\citenamefont {Borges}\ \emph {et~al.}(2018)\citenamefont {Borges},
  \citenamefont {Rossatto}, \citenamefont {Luiz},\ and\ \citenamefont
  {Villas-Boas}}]{cvbpra18}%
  \BibitemOpen
  \bibfield  {author} {\bibinfo {author} {\bibfnamefont {H.~S.}\ \bibnamefont
  {Borges}}, \bibinfo {author} {\bibfnamefont {D.~Z.}\ \bibnamefont
  {Rossatto}}, \bibinfo {author} {\bibfnamefont {F.~S.}\ \bibnamefont {Luiz}},\
  and\ \bibinfo {author} {\bibfnamefont {C.~J.}\ \bibnamefont {Villas-Boas}},\
  }\bibfield  {title} {\bibinfo {title} {Heralded entangling quantum gate via
  cavity-assisted photon scattering},\ }\href
  {https://doi.org/10.1103/PhysRevA.97.013828} {\bibfield  {journal} {\bibinfo
  {journal} {Phys. Rev. A}\ }\textbf {\bibinfo {volume} {97}},\ \bibinfo
  {pages} {013828} (\bibinfo {year} {2018})}\BibitemShut {NoStop}%
\bibitem [{\citenamefont {Hacker}\ \emph {et~al.}(2016)\citenamefont {Hacker},
  \citenamefont {Welte}, \citenamefont {Rempe},\ and\ \citenamefont
  {Ritter}}]{Hacker2016}%
  \BibitemOpen
  \bibfield  {author} {\bibinfo {author} {\bibfnamefont {B.}~\bibnamefont
  {Hacker}}, \bibinfo {author} {\bibfnamefont {S.}~\bibnamefont {Welte}},
  \bibinfo {author} {\bibfnamefont {G.}~\bibnamefont {Rempe}},\ and\ \bibinfo
  {author} {\bibfnamefont {S.}~\bibnamefont {Ritter}},\ }\bibfield  {title}
  {\bibinfo {title} {A photon{\textendash}photon quantum gate based on a single
  atom in an optical resonator},\ }\href {https://doi.org/10.1038/nature18592}
  {\bibfield  {journal} {\bibinfo  {journal} {Nature (London)}\ }\textbf
  {\bibinfo {volume} {536}},\ \bibinfo {pages} {193} (\bibinfo {year}
  {2016})}\BibitemShut {NoStop}%
\bibitem [{\citenamefont {Daiss}\ \emph {et~al.}(2021)\citenamefont {Daiss},
  \citenamefont {Langenfeld}, \citenamefont {Welte}, \citenamefont {Distante},
  \citenamefont {Thomas}, \citenamefont {Hartung}, \citenamefont {Morin},\ and\
  \citenamefont {Rempe}}]{Daiss2021}%
  \BibitemOpen
  \bibfield  {author} {\bibinfo {author} {\bibfnamefont {S.}~\bibnamefont
  {Daiss}}, \bibinfo {author} {\bibfnamefont {S.}~\bibnamefont {Langenfeld}},
  \bibinfo {author} {\bibfnamefont {S.}~\bibnamefont {Welte}}, \bibinfo
  {author} {\bibfnamefont {E.}~\bibnamefont {Distante}}, \bibinfo {author}
  {\bibfnamefont {P.}~\bibnamefont {Thomas}}, \bibinfo {author} {\bibfnamefont
  {L.}~\bibnamefont {Hartung}}, \bibinfo {author} {\bibfnamefont
  {O.}~\bibnamefont {Morin}},\ and\ \bibinfo {author} {\bibfnamefont
  {G.}~\bibnamefont {Rempe}},\ }\bibfield  {title} {\bibinfo {title} {A
  quantum-logic gate between distant quantum-network modules},\ }\href
  {https://doi.org/10.1126/science.abe3150} {\bibfield  {journal} {\bibinfo
  {journal} {Science}\ }\textbf {\bibinfo {volume} {371}},\ \bibinfo {pages}
  {614} (\bibinfo {year} {2021})}\BibitemShut {NoStop}%
\bibitem [{\citenamefont {Stolz}\ \emph {et~al.}(2022)\citenamefont {Stolz},
  \citenamefont {Hegels}, \citenamefont {Winter}, \citenamefont {R\"ohr},
  \citenamefont {Hsiao}, \citenamefont {Husel}, \citenamefont {Rempe},\ and\
  \citenamefont {D\"urr}}]{PhysRevX.12.021035}%
  \BibitemOpen
  \bibfield  {author} {\bibinfo {author} {\bibfnamefont {T.}~\bibnamefont
  {Stolz}}, \bibinfo {author} {\bibfnamefont {H.}~\bibnamefont {Hegels}},
  \bibinfo {author} {\bibfnamefont {M.}~\bibnamefont {Winter}}, \bibinfo
  {author} {\bibfnamefont {B.}~\bibnamefont {R\"ohr}}, \bibinfo {author}
  {\bibfnamefont {Y.-F.}\ \bibnamefont {Hsiao}}, \bibinfo {author}
  {\bibfnamefont {L.}~\bibnamefont {Husel}}, \bibinfo {author} {\bibfnamefont
  {G.}~\bibnamefont {Rempe}},\ and\ \bibinfo {author} {\bibfnamefont
  {S.}~\bibnamefont {D\"urr}},\ }\bibfield  {title} {\bibinfo {title}
  {{Quantum-Logic Gate between Two Optical Photons with an Average Efficiency
  above 40\%}},\ }\href {https://doi.org/10.1103/PhysRevX.12.021035} {\bibfield
   {journal} {\bibinfo  {journal} {Phys. Rev. X}\ }\textbf {\bibinfo {volume}
  {12}},\ \bibinfo {pages} {021035} (\bibinfo {year} {2022})}\BibitemShut
  {NoStop}%
\bibitem [{\citenamefont {Hofmann}\ \emph {et~al.}(2003)\citenamefont
  {Hofmann}, \citenamefont {Kojima}, \citenamefont {Takeuchi},\ and\
  \citenamefont {Sasaki}}]{Holger2003}%
  \BibitemOpen
  \bibfield  {author} {\bibinfo {author} {\bibfnamefont {H.~F.}\ \bibnamefont
  {Hofmann}}, \bibinfo {author} {\bibfnamefont {K.}~\bibnamefont {Kojima}},
  \bibinfo {author} {\bibfnamefont {S.}~\bibnamefont {Takeuchi}},\ and\
  \bibinfo {author} {\bibfnamefont {K.}~\bibnamefont {Sasaki}},\ }\bibfield
  {title} {\bibinfo {title} {Optimized phase switching using a single-atom
  nonlinearity},\ }\href {https://dx.doi.org/10.1088/1464-4266/5/3/304}
  {\bibfield  {journal} {\bibinfo  {journal} {J. Opt. B: Quantum Semiclassical
  Opt.}\ }\textbf {\bibinfo {volume} {5}},\ \bibinfo {pages} {218} (\bibinfo
  {year} {2003})}\BibitemShut {NoStop}%
\bibitem [{\citenamefont {Gardiner}\ and\ \citenamefont
  {Zoller}(2010)}]{qnoise}%
  \BibitemOpen
  \bibfield  {author} {\bibinfo {author} {\bibfnamefont {C.}~\bibnamefont
  {Gardiner}}\ and\ \bibinfo {author} {\bibfnamefont {P.}~\bibnamefont
  {Zoller}},\ }\href {https://books.google.com.br/books?id=PygdvgAACAAJ} {\emph
  {\bibinfo {title} {Quantum Noise: A Handbook of Markovian and Non-Markovian
  Quantum Stochastic Methods with Applications to Quantum Optics}}}\ (\bibinfo
  {publisher} {Springer-Verlag, Berlin},\ \bibinfo {year} {2010})\BibitemShut
  {NoStop}%
\bibitem [{\citenamefont {Loudon}(2000)}]{Loudon2000}%
  \BibitemOpen
  \bibfield  {author} {\bibinfo {author} {\bibfnamefont {R.}~\bibnamefont
  {Loudon}},\ }\href {https://books.google.com.br/books?id=AEkfajgqldoC} {\emph
  {\bibinfo {title} {The Quantum Theory of Light}}}\ (\bibinfo  {publisher}
  {Oxford University Press, Oxford},\ \bibinfo {year} {2000})\BibitemShut
  {NoStop}%
\bibitem [{\citenamefont {Walls}\ and\ \citenamefont
  {Milburn}(2008)}]{walls2008}%
  \BibitemOpen
  \bibfield  {author} {\bibinfo {author} {\bibfnamefont {D.}~\bibnamefont
  {Walls}}\ and\ \bibinfo {author} {\bibfnamefont {G.}~\bibnamefont
  {Milburn}},\ }\href {https://books.google.com.br/books?id=LiWsc3Nlf0kC}
  {\emph {\bibinfo {title} {Quantum Optics}}}\ (\bibinfo  {publisher} {Springer
  Berlin Heidelberg},\ \bibinfo {year} {2008})\BibitemShut {NoStop}%
\bibitem [{\citenamefont {Primakoff}\ and\ \citenamefont
  {Holstein}(1939)}]{holsteinprimakoff}%
  \BibitemOpen
  \bibfield  {author} {\bibinfo {author} {\bibfnamefont {H.}~\bibnamefont
  {Primakoff}}\ and\ \bibinfo {author} {\bibfnamefont {T.}~\bibnamefont
  {Holstein}},\ }\bibfield  {title} {\bibinfo {title} {{Many-Body Interactions
  in Atomic and Nuclear Systems}},\ }\href
  {https://doi.org/10.1103/PhysRev.55.1218} {\bibfield  {journal} {\bibinfo
  {journal} {Phys. Rev.}\ }\textbf {\bibinfo {volume} {55}},\ \bibinfo {pages}
  {1218} (\bibinfo {year} {1939})}\BibitemShut {NoStop}%
\bibitem [{\citenamefont {Bonifacio}\ and\ \citenamefont
  {Preparata}(1970)}]{PhysRevA.2.336}%
  \BibitemOpen
  \bibfield  {author} {\bibinfo {author} {\bibfnamefont {R.}~\bibnamefont
  {Bonifacio}}\ and\ \bibinfo {author} {\bibfnamefont {G.}~\bibnamefont
  {Preparata}},\ }\bibfield  {title} {\bibinfo {title} {Coherent {S}pontaneous
  {E}mission},\ }\href {https://doi.org/10.1103/PhysRevA.2.336} {\bibfield
  {journal} {\bibinfo  {journal} {Phys. Rev. A}\ }\textbf {\bibinfo {volume}
  {2}},\ \bibinfo {pages} {336} (\bibinfo {year} {1970})}\BibitemShut {NoStop}%
\bibitem [{\citenamefont {Dilley}\ \emph {et~al.}(2012)\citenamefont {Dilley},
  \citenamefont {Nisbet-Jones}, \citenamefont {Shore},\ and\ \citenamefont
  {Kuhn}}]{kuhn2012}%
  \BibitemOpen
  \bibfield  {author} {\bibinfo {author} {\bibfnamefont {J.}~\bibnamefont
  {Dilley}}, \bibinfo {author} {\bibfnamefont {P.}~\bibnamefont
  {Nisbet-Jones}}, \bibinfo {author} {\bibfnamefont {B.~W.}\ \bibnamefont
  {Shore}},\ and\ \bibinfo {author} {\bibfnamefont {A.}~\bibnamefont {Kuhn}},\
  }\bibfield  {title} {\bibinfo {title} {Single-photon absorption in coupled
  atom-cavity systems},\ }\href {https://doi.org/10.1103/PhysRevA.85.023834}
  {\bibfield  {journal} {\bibinfo  {journal} {Phys. Rev. A}\ }\textbf {\bibinfo
  {volume} {85}},\ \bibinfo {pages} {023834} (\bibinfo {year}
  {2012})}\BibitemShut {NoStop}%
\bibitem [{\citenamefont {Baragiola}\ \emph {et~al.}(2012)\citenamefont
  {Baragiola}, \citenamefont {Cook}, \citenamefont {Bra\ifmmode~\acute{n}\else
  \'{n}\fi{}czyk},\ and\ \citenamefont {Combes}}]{baragiola}%
  \BibitemOpen
  \bibfield  {author} {\bibinfo {author} {\bibfnamefont {B.~Q.}\ \bibnamefont
  {Baragiola}}, \bibinfo {author} {\bibfnamefont {R.~L.}\ \bibnamefont {Cook}},
  \bibinfo {author} {\bibfnamefont {A.~M.}\ \bibnamefont
  {Bra\ifmmode~\acute{n}\else \'{n}\fi{}czyk}},\ and\ \bibinfo {author}
  {\bibfnamefont {J.}~\bibnamefont {Combes}},\ }\bibfield  {title} {\bibinfo
  {title} {{$N$}-photon wave packets interacting with an arbitrary quantum
  system},\ }\href {https://doi.org/10.1103/PhysRevA.86.013811} {\bibfield
  {journal} {\bibinfo  {journal} {Phys. Rev. A}\ }\textbf {\bibinfo {volume}
  {86}},\ \bibinfo {pages} {013811} (\bibinfo {year} {2012})}\BibitemShut
  {NoStop}%
\bibitem [{\citenamefont {Barnett}\ \emph {et~al.}(1998)\citenamefont
  {Barnett}, \citenamefont {Jeffers}, \citenamefont {Gatti},\ and\
  \citenamefont {Loudon}}]{PhysRevA.57.2134}%
  \BibitemOpen
  \bibfield  {author} {\bibinfo {author} {\bibfnamefont {S.~M.}\ \bibnamefont
  {Barnett}}, \bibinfo {author} {\bibfnamefont {J.}~\bibnamefont {Jeffers}},
  \bibinfo {author} {\bibfnamefont {A.}~\bibnamefont {Gatti}},\ and\ \bibinfo
  {author} {\bibfnamefont {R.}~\bibnamefont {Loudon}},\ }\bibfield  {title}
  {\bibinfo {title} {Quantum optics of lossy beam splitters},\ }\href
  {https://doi.org/10.1103/PhysRevA.57.2134} {\bibfield  {journal} {\bibinfo
  {journal} {Phys. Rev. A}\ }\textbf {\bibinfo {volume} {57}},\ \bibinfo
  {pages} {2134} (\bibinfo {year} {1998})}\BibitemShut {NoStop}%
\bibitem [{\citenamefont {Delanty}\ \emph {et~al.}(2011)\citenamefont
  {Delanty}, \citenamefont {Rebic},\ and\ \citenamefont
  {Twamley}}]{Delanty2011}%
  \BibitemOpen
  \bibfield  {author} {\bibinfo {author} {\bibfnamefont {M.}~\bibnamefont
  {Delanty}}, \bibinfo {author} {\bibfnamefont {S.}~\bibnamefont {Rebic}},\
  and\ \bibinfo {author} {\bibfnamefont {J.}~\bibnamefont {Twamley}},\
  }\href@noop {} {\bibinfo {title} {Superradiance of harmonic oscillators}}
  (\bibinfo {year} {2011}),\ \Eprint {https://arxiv.org/abs/arXiv:1107.5080}
  {arXiv:1107.5080} \BibitemShut {NoStop}%
\bibitem [{\citenamefont {Máximo}\ \emph {et~al.}(2021)\citenamefont
  {Máximo}, \citenamefont {de~Souza}, \citenamefont {Ianzano}, \citenamefont
  {Rempe}, \citenamefont {Bachelard},\ and\ \citenamefont
  {Villas-Boas}}]{Mximo2021}%
  \BibitemOpen
  \bibfield  {author} {\bibinfo {author} {\bibfnamefont {C.~E.}\ \bibnamefont
  {Máximo}}, \bibinfo {author} {\bibfnamefont {P.~P.}\ \bibnamefont
  {de~Souza}}, \bibinfo {author} {\bibfnamefont {C.}~\bibnamefont {Ianzano}},
  \bibinfo {author} {\bibfnamefont {G.}~\bibnamefont {Rempe}}, \bibinfo
  {author} {\bibfnamefont {R.}~\bibnamefont {Bachelard}},\ and\ \bibinfo
  {author} {\bibfnamefont {C.~J.}\ \bibnamefont {Villas-Boas}},\ }\href@noop {}
  {\bibinfo {title} {{Bright and Dark States of Light: The Quantum Origin of
  Classical Interference}}} (\bibinfo {year} {2021}),\ \Eprint
  {https://arxiv.org/abs/arXiv:2112.05512} {arXiv:2112.05512} \BibitemShut
  {NoStop}%
\bibitem [{\citenamefont {Autler}\ and\ \citenamefont
  {Townes}(1955)}]{autlertownes}%
  \BibitemOpen
  \bibfield  {author} {\bibinfo {author} {\bibfnamefont {S.~H.}\ \bibnamefont
  {Autler}}\ and\ \bibinfo {author} {\bibfnamefont {C.~H.}\ \bibnamefont
  {Townes}},\ }\bibfield  {title} {\bibinfo {title} {{Stark Effect in Rapidly
  Varying Fields}},\ }\href {https://doi.org/10.1103/PhysRev.100.703}
  {\bibfield  {journal} {\bibinfo  {journal} {Phys. Rev.}\ }\textbf {\bibinfo
  {volume} {100}},\ \bibinfo {pages} {703} (\bibinfo {year}
  {1955})}\BibitemShut {NoStop}%
\bibitem [{\citenamefont {Fleischhauer}\ \emph {et~al.}(2005)\citenamefont
  {Fleischhauer}, \citenamefont {Imamoglu},\ and\ \citenamefont
  {Marangos}}]{fleisch}%
  \BibitemOpen
  \bibfield  {author} {\bibinfo {author} {\bibfnamefont {M.}~\bibnamefont
  {Fleischhauer}}, \bibinfo {author} {\bibfnamefont {A.}~\bibnamefont
  {Imamoglu}},\ and\ \bibinfo {author} {\bibfnamefont {J.~P.}\ \bibnamefont
  {Marangos}},\ }\bibfield  {title} {\bibinfo {title} {Electromagnetically
  induced transparency: {Optics} in coherent media},\ }\href
  {https://doi.org/10.1103/RevModPhys.77.633} {\bibfield  {journal} {\bibinfo
  {journal} {Rev. Mod. Phys.}\ }\textbf {\bibinfo {volume} {77}},\ \bibinfo
  {pages} {633} (\bibinfo {year} {2005})}\BibitemShut {NoStop}%
\bibitem [{\citenamefont {Alsing}\ \emph {et~al.}(1992)\citenamefont {Alsing},
  \citenamefont {Cardimona},\ and\ \citenamefont {Carmichael}}]{Alsing}%
  \BibitemOpen
  \bibfield  {author} {\bibinfo {author} {\bibfnamefont {P.~M.}\ \bibnamefont
  {Alsing}}, \bibinfo {author} {\bibfnamefont {D.~A.}\ \bibnamefont
  {Cardimona}},\ and\ \bibinfo {author} {\bibfnamefont {H.~J.}\ \bibnamefont
  {Carmichael}},\ }\bibfield  {title} {\bibinfo {title} {Suppression of
  fluorescence in a lossless cavity},\ }\href
  {https://doi.org/10.1103/PhysRevA.45.1793} {\bibfield  {journal} {\bibinfo
  {journal} {Phys. Rev. A}\ }\textbf {\bibinfo {volume} {45}},\ \bibinfo
  {pages} {1793} (\bibinfo {year} {1992})}\BibitemShut {NoStop}%
\bibitem [{\citenamefont {Rice}\ and\ \citenamefont {Brecha}(1996)}]{Rice1996}%
  \BibitemOpen
  \bibfield  {author} {\bibinfo {author} {\bibfnamefont {P.}~\bibnamefont
  {Rice}}\ and\ \bibinfo {author} {\bibfnamefont {R.}~\bibnamefont {Brecha}},\
  }\bibfield  {title} {\bibinfo {title} {Cavity induced transparency},\ }\href
  {https://doi.org/10.1016/0030-4018(96)00102-2} {\bibfield  {journal}
  {\bibinfo  {journal} {Opt. Commun.}\ }\textbf {\bibinfo {volume} {126}},\
  \bibinfo {pages} {230} (\bibinfo {year} {1996})}\BibitemShut {NoStop}%
\bibitem [{spr()}]{spread2}%
  \BibitemOpen
  \href@noop {} {\bibinfo {title} {Considering the atom as a control qubit, it
  will be able to perform logical operations on a string of pulses of light
  and, eventually, we will have an entangled state among the atom and all the
  pulses that impinge upon the crossed-cavity system. {O}n the other hand,
  considering several atoms inside the cavities, as target qubits, and the
  cavity modes as the control qubit, the atomic state could acquire a phase
  shift depending on the collective state (dark or bright) of the modes, thus
  implementing a multi-qubit gate in a single shot.}}\BibitemShut {Stop}%
\bibitem [{\citenamefont {Guerra~Bobo}(2013)}]{GuerraBobo2013}%
  \BibitemOpen
  \bibfield  {author} {\bibinfo {author} {\bibfnamefont {I.}~\bibnamefont
  {Guerra~Bobo}},\ }\bibfield  {title} {\bibinfo {title} {On quantum
  conditional probability},\ }\href {https://doi.org/10.1387/theoria.5682}
  {\bibfield  {journal} {\bibinfo  {journal} {THEORIA}\ }\textbf {\bibinfo
  {volume} {28}},\ \bibinfo {pages} {115–137} (\bibinfo {year}
  {2013})}\BibitemShut {NoStop}%
\end{thebibliography}%

\end{document}